\documentclass[aps,prd,preprint,superscriptaddress,amsmath,amssymb,showpacs]{revtex4-1}
\usepackage{dcolumn}
\usepackage{graphicx}
\usepackage{float}
\usepackage{physics}
\usepackage{amsmath}
\usepackage{ulem}
\usepackage{xcolor}
\begin{document}

\title{The Interaction of Moving $\mathbf{Q\bar{Q}}$ and QQq in the Thermal Plasma}

\author{Xuan Liu}
\affiliation{School of Nuclear Science and Technology, University of South China, Hengyang 421001, China}
\author{Sheng Lin}
\affiliation{School of Nuclear Science and Technology, University of South China, Hengyang 421001, China}
\author{Xun Chen}
\email{chenxun@usc.edu.cn}
\affiliation{School of Nuclear Science and Technology, University of South China, Hengyang 421001, China}
\affiliation{Key Laboratory of Quark and Lepton Physics (MOE), Central China Normal University, Wuhan 430079,China}
\affiliation{INFN -- Istituto Nazionale di Fisica Nucleare -- Sezione di Bari, Via Orabona 4, 70125 Bari, Italy}
\begin{abstract}
    The strength of the interaction between heavy quarks is studied for heavy quarkonium ($\mathrm{Q\Bar{Q}}$) and doubly heavy baryons ($\mathrm{QQq}$) at finite temperature and rapidity using the gauge/gravity duality in this paper. We show that this theoretical framework is capable of simultaneously and accurately describing both $\mathrm{Q\Bar{Q}}$ and $\mathrm{QQq}$ by fitting lattice potentials. In this framework, we study their interaction at long distances or low temperature and rapidity through effective string tension, while the interaction at short distances or high temperature and rapidity is studied through effective running coupling. Additionally, we plot their state diagram in the $T-\eta$ plane and systematically calculate their respective screening distances.

\end{abstract}

\maketitle

\section{Introduction}
The interaction between quarks is essential for our understanding of $\mathrm{QCD}$ and the $\mathrm{QCD}$ phase transition. When quarks are confined within hadrons, the force between them increases with the separation distance until the string breaks, resulting in the creation of new quark-antiquark pairs, thereby preventing the isolation of free quarks. $\mathrm{QCD}$ predicts that deconfinement occurs under extreme conditions, where color screening happens at long distances between quarks, leading to the emergence of relatively free quarks \cite{Matsui:1986dk, Kharzeev:1995kz, Witten:1998zw}. A system composed of such deconfined particles is called a Quark-Gluon Plasma ($\mathrm{QGP}$) \cite{Gyulassy:2004zy, Jacobs:2004qv}. Lattice QCD defined an effective running coupling \cite{Kaczmarek:2004gv, Kaczmarek:2005ui,Kaczmarek:2007pb,Bazavov:2018wmo},
\begin{equation}\label{01}
\alpha_{Q \bar{Q}}(r)=\frac{1}{C_F} r^2 \frac{\partial E(r)}{\partial r},
\end{equation}
to study the force between the static quark and antiquark \cite{Gubser:2006qh}. $C_F=4 / 3$ is the Casimir operator in the fundamental representation of $\mathrm{SU}(3)$ and $E$ is the static energy of $\mathrm{Q\Bar{Q}}$. In addition, the running coupling constant, although defined differently in this work, plays a key role in many important physical processes, such as quark confinement and hadronization \cite{CarcamoHernandez:2013ydh, ParticleDataGroup:2024cfk, Deur:2016tte}.

Creating $\mathrm{QGP}$ through heavy-ion collisions recreates the extreme conditions of high temperature and rapid expansion from the early universe \cite{Laine:2006we, Asaka:2006rw, Hindmarsh:2005ix, Rothkopf:2019ipj, Shuryak:2014zxa, 2015The, Busza:2018rrf}. This helps us to understand the evolution of the early universe and the properties of $\mathrm{QCD}$. Heavy quarkonium is an important probe for studying conditions of extreme high temperature and rapid expansion \cite{QuarkoniumWorkingGroup:2004kpm, Brambilla:2010cs, vanHees:2005wb}. Moreover, in the recent $\mathrm{LHCb}$ experiment at $\mathrm{CERN}$, researchers have discovered a new particle known as $\Xi_{cc}^{++}$ \cite{LHCb:2017iph, LHCb:2018pcs}. It is composed of two heavy quarks and one light quark, and its discovery has greatly increased interest in the study of doubly heavy baryon. While the running coupling constant has been extensively studied from many different approaches, see e.g. Refs. \cite{Giunti:1991ta, Baikov:2016tgj, Javidan:2020lup, Aguilar:2001zy, Lombardo:1996gp, Zhou:2023qtr, Bloch:2003sk, Deur:2016tte, Yu:2021yvw, Takaura:2018vcy, Chen:2021gop}, here we use the definition given in Refs. \cite{Kaczmarek:2005ui,Kaczmarek:2007pb,Bazavov:2018wmo, Kaczmarek:2004gv, Kaczmarek:2008saj} as in Eq. (\ref{01}). It is well known that the $\mathrm{QGP}$ expands rapidly after its formation; however, during its cooling and expansion, heavy quarks with finite velocity relative to the QGP are invariably produced. Therefore, relative rapidity is an unavoidable factor to consider when discussing the interaction forces of particles. Taking the Au+Au collisions at RHIC as an example, various particle species emerge during the diffusion and cooling processes of the $\mathrm{QGP}$ \cite{STAR:2021orx, Jin:2024xxz, STAR:2024znc}. Therefore, the consistent description of different particle behaviors under the same physical framework carries significant physical implications. This paper attempts to reveal more information about the $\mathrm{QGP}$ by contrasting heavy quarkonium and doubly heavy baryon at the same physical framework. Moreover, the interaction forces between quarks in the $\mathrm{QGP}$ can also reflect the state of the $\mathrm{QGP}$ to a certain extent, such as determining whether it is a strongly coupled $\mathrm{QGP}$ ($\mathrm{sQGP}$) or a weakly coupled $\mathrm{QGP}$ ($\mathrm{wQGP}$) \cite{Nijs:2023bok, Shuryak:2014zxa, Arnold:2004ti, Shuryak:2004cy}.

Lattice gauge theory remains the fundamental tool for studying non-perturbative phenomena in $\mathrm{QCD}$, yet its application to doubly heavy baryon has been relatively limited \cite{Yamamoto:2008jz, Najjar:2009da}. Gauge/gravity duality offers a new avenue for probing strongly coupled gauge theories. Originally,  Maldacena \cite{Maldacena:1997re} proposed the gauge/gravity duality for conformal field theories, but it was subsequently extended to encompass theories akin to $\mathrm{QCD}$, thereby establishing to some extent a linkage between string theory and heavy-ion collisions \cite{Aharony:1999ti, Casalderrey-Solana:2011dxg, Gubser:2011qv}. In recent years, there has been extensive research on moving heavy quarkonium. A computational framework based on effective field theory and open quantum systems has been systematically established and closely integrated with lattice QCD calculations \cite{Brambilla:2016wgg, Akamatsu:2020ypb, Yao:2021lus, Nijs:2023dbc, Brambilla:2024tqg, Brambilla:2025cqy, Daddi-Hammou:2025hdz}, while holographic QCD offers a novel perspective on these studies \cite{Liu:2006nn, Chen:2020ath, Kajantie:2006hv, Andreev:2006eh, Chen:2017lsf, Finazzo:2014rca, Ali-Akbari:2014vpa, Andreev:2021vjr, Krishnan:2008rs, Chernicoff:2012bu, BitaghsirFadafan:2015yng, Zhou:2020ssi, Feng:2019boe, Zhou:2021sdy}, demonstrating consistent findings with Lattice QCD results \cite{Zeng:2008sx, Chen:2025kqb, Luo:2024iwf, Chen:2024mmd, Chen:2024ckb}.
Moreover, the multi-quark potential obtained through effective string model is in good agreement with lattice results \cite{Alexandrou:2001yt, Alexandrou:2002sn, Takahashi:2002bw, Andreev:2020xor, Andreev:2015iaa, Andreev:2015riv, Andreev:2019cbc, Andreev:2021bfg, Andreev:2022qdu, Andreev:2022cax, Andreev:2023hmh, Mei:2022dkd, Andreev:2021eyj, Mei:2024rjg}.

This paper shows a unified framework to properly describe the behaviors of both heavy quarkonium and doubly heavy baryons in a QGP at finite temperature and rapidity, ensuring consistency with lattice QCD results while providing a detailed discussion of the interparticle interaction forces within this framework. The rest of the paper is organized as follows: in the Sec. \ref{sec:2}, we present the theoretical framework and show a unified description for both particles by fitting their lattice potentials. In Sec. \ref{sec:3}, we introduce the string tension within this unified framework to discuss and validate the QCD phase transition. Based on this approach, we plot the state diagrams of the two types of particles in the $T-\eta$ plane. Subsequently, we conduct a detailed discussion on the interactions of heavy quarkonium and doubly heavy baryons through the effective coupling, including: the effective running coupling with distance, and with temperature and rapidity, as well as the impact of rapidity on the temperature dependence of the effective coupling constant. Additionally, we examine the effects of temperature and rapidity on their screening distances. In the Sec. \ref{sec.4}, we provide a summary of this paper.

\section{Preliminaries}\label{sec:2}
The effective string holographic models we study in this work have been proposed by Andreev recently. These models can not only describe the potential of heavy quarkonium \cite{Andreev:2006nw}, but also describe the potential of exotic hadrons \cite{Andreev:2015riv,Andreev:2015iaa, Andreev:2007rx,Andreev:2024orz,Andreev:2022qdu,Andreev:2023hmh,Chen:2021bkc}. The purpose of this study is to reveal the properties of $\mathrm{Q\Bar{Q}}$ and $\mathrm{QQq}$ based on the effective string model by investigating the effective running coupling and the screening distance of their motion in a thermal medium. First, we present the metric \cite{Avramis:2006em, Caceres:2006ta, Natsuume:2007vc, Liu:2008tz}
\begin{align}
    ds^{2}&=w(r) \Big (-f(r)dt^{2}+d\Vec{x}^{2}+\frac{1}{f(r)}dr^{2} \Big )+e^{-\mathbf{s}r^{2}}g_{ab}^{(5)}d\omega^{a}d\omega^{b},\label{1}
\end{align}
where
\begin{equation}
    \begin{aligned}
        w(r)&=\frac{e^{\mathbf{s}r^{2}}R^{2}}{r^{2}}\,,\\ f(r)&=1-\frac{r^{4}}{r_{h}^{4}}.\label{2}
    \end{aligned}
\end{equation}

The metric signifies a deformation of the Euclidean $\mathrm{AdS_5} \times \mathrm{S_5}$ space, controlled by a single parameter $\mathbf{s}$ and with a radius $R = 1$. In this work, $\mathbf{s}$ is determined to be $0.41\,\mathrm{GeV^2}$ by fitting the lattice potential of heavy quarkonium. Therefore, the metric is composed of an $\mathrm{AdS_5}$ space and a five-dimensional compact space $\mathrm {X}$ with coordinates $\omega^{a}$. The function $f(r)$ is the blackening factor, which decreases within the interval $[0, r_h]$, where $r_h$ represents the black hole horizon (brane). Additionally, when hadrons are confined, an imaginary wall exists at $r_w$ on the $r$-axis \cite{Karch:2006pv, Wen:2024hgu, Liang:2023lgs, Cao:2022csq, Cao:2022mep, Yang:2015aia, Andreev:2006nw, Colangelo:2010pe}. The Hawking temperature $T$ associated with the black hole is given by:
\begin{equation}
    T=\frac{1}{4\pi}{\left | \frac{df}{dr}  \right |}_{r=r_{h}}=\frac{1}{\pi r_{h}}.\label{3}
\end{equation}

In this paper, we consider a $\mathrm{Q\bar{Q}}$ or QQq moving with rapidity $\eta$ through a thermal medium at temperature $T$. We can assume that the $\mathrm{Q\bar{Q}}$ or QQq is at rest, while the thermal medium moves relative to it at rapidity $\eta$, which can be considered as a "thermal wind" blowing past the $\mathrm{Q\bar{Q}}$ or QQq in the $x_3$ direction \cite{Liu:2006nn, Finazzo:2014rca, Chen:2017lsf, Andreev:2021vjr, Thakur:2016cki}. Through a Lorentz transformation, we can provide a new background metric:
\begin{align}
    ds^{2}=w(r)&\Big(-g_{1}(r)dt^{2}
 -2\sinh(\eta)\cosh(\eta)\big(1-\frac{g_{1}(r)}{g_{2}(r)}\big)dx_{3}dt\notag\\ &+g_{3}(r)dx_{3}^{2}+dx_{1}^{2}+dx_{2}^{2}+\frac{g_{2}(r)}{g_{1}(r)}dr^{2}\Big)+e^{-\mathbf{s}r^{2}}g_{ab}^{(5)}d\omega^{a}d\omega^{b}, \label{4}
\end{align}
where
\begin{equation}
    \begin{aligned}
        g_{1}(r)&=f(r)\cosh^{2}(\eta)-\sinh^{2}(\eta),\\g_{2}(r)&=\frac{g_{1}(r)}{f(r)},\\g_{3}(r)&=\cosh^{2}(\eta)-f(r)\sinh^{2}(\eta).\label{5}
    \end{aligned}
\end{equation}

The Nambu-Goto action of a string is
\begin{equation}
     S_{NG} = - g \int d\xi^{0} d\xi^{1} \sqrt{- \det g_{ab}},\label{6}
\end{equation}
where the $g_{ab}$ is an induced metric ($\xi^{0}$, $\xi^{1}$) are worldsheet coordinates, and $g$ is related to the string tension.
Based on the $\mathrm{AdS/CFT}$ correspondence, we know that the baryon vertex corresponds to a five brane \cite{Witten:1997bs, Gukov:1998kn}, the vertex action is $S_{vert} = \tau_{5} \int d^{6}\xi \sqrt{\gamma^{(6)}}$, where $\tau_{5}$ is the brane tension and $\xi^{i}$ are the world-volume coordinates. Since the brane is wrapped on the compact space $\mathrm {X}$, it appears point-like in $\mathrm{AdS_5}$. As will be discussed in subsequent sections, the position $r_v$ of the baryon vertex determines the configurational stage of the doubly heavy baryon. We choose a static gauge where $\xi^{0} = t$ and $\xi^{a} = \theta^{a}$, with $\theta^{a}$ being the coordinates on $\mathrm {X}$. Consequently, the action is:
\begin{equation}
    S_{vert}=\tau_{v}\int dt\frac{e^{-2\mathbf{s}r^2}}{r} \sqrt{g_{1}(r)},\label{7}
\end{equation}
where $\tau_{v}$ is a dimensionless parameter defined by $\tau_{v}=\mathrm{\tau}_{5}R\mathrm{vol(X)}$ and $\mathrm{vol(X)}$ is a volume of $\mathrm {X}$. Finally, we consider the light quarks at the endpoints of the string as a tachyon field, which is coupled to the worldsheet boundary through $S_q=\int d\tau \mathbf{e}\mathrm{T}$, where $\mathrm{T}(x, r)$ is a scalar field that describes open string tachyon, $\tau$ is a coordinate on the boundary, and $\mathbf{e}$ is the boundary metric \cite{Andreev:2020pqy, Erlich:2005qh}. We consider only the case where $\mathrm{T}(x, r)=m$ and the worldsheet boundary is a line in the $t$ direction, in which case the action can be written as:
\begin{equation}
    S_{q}=m\int dt\frac{e^{\frac{\mathbf{s}r^2}{2}}}{r} \sqrt{g_{1}(r)},
\end{equation}
This action represents a particle of mass $m$ at rest, with a medium at temperature $T$ moving past it at a rapidity $\eta$.

\subsection{Heavy quarkonium}\label{A}

In this paper, we investigate the scenario where the direction of motion is perpendicular to the heavy quark pair, with the heavy quark pair located at $x_1$ and the rapidity $\eta$ along the $x_3$ direction. For brevity, $x$ will be used to denote $x_1$ in the following. Then we choose the static gauge $\xi^{0}=t$, $\xi^{1}=x$, and the action of $\mathrm{Q\Bar{Q}}$ can be written as:
\begin{align}
     S_{\mathrm{Q\Bar{Q}}}&=g_{\mathrm{Q\Bar{Q}}}\int_{0}^{t}\int_{-\frac{L}{2}}^{\frac{L}{2}} w(r)\sqrt{g_{1}(r)+g_{2}(r)(\partial_{x}r)^2} dx \\&=g_{\mathrm{Q\Bar{Q}}}t\int_{-\frac{L}{2}}^{\frac{L}{2}} w(r)\sqrt{g_{1}(r)+g_{2}(r)(\partial_{x}r)^2} dx.
\end{align}

\begin{figure}
    \centering
    \includegraphics[width=8.5cm]{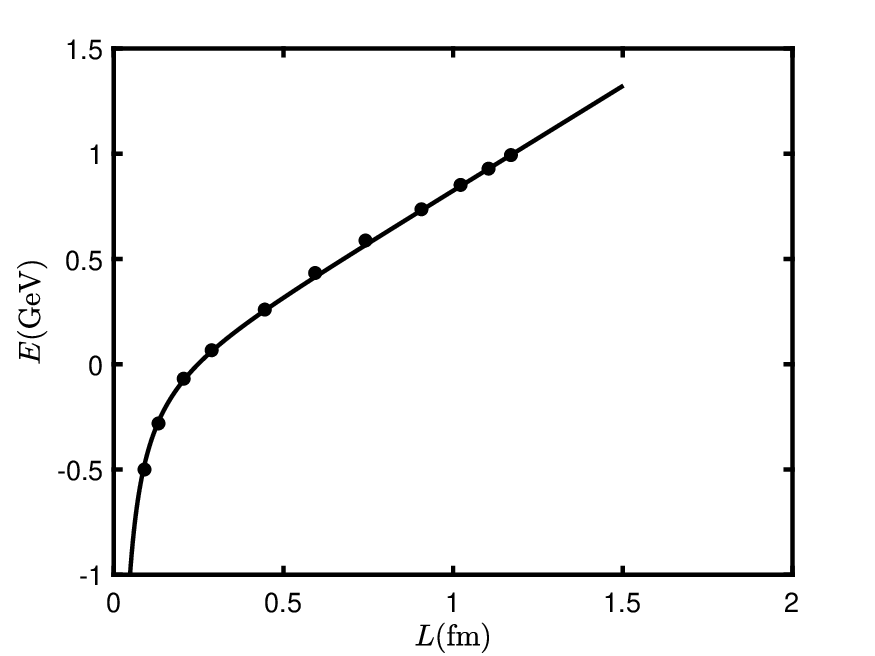}
    \caption{\label{fig1} The black curve represents the $\mathrm{Q\Bar{Q}}$ potential at $T=0, \eta =0$. The black solid dot represents the lattice data from quenched QCD \cite{Kaczmarek:2004gv}.}
\end{figure}

\begin{figure}
    \centering
    \includegraphics[width=8.5cm]{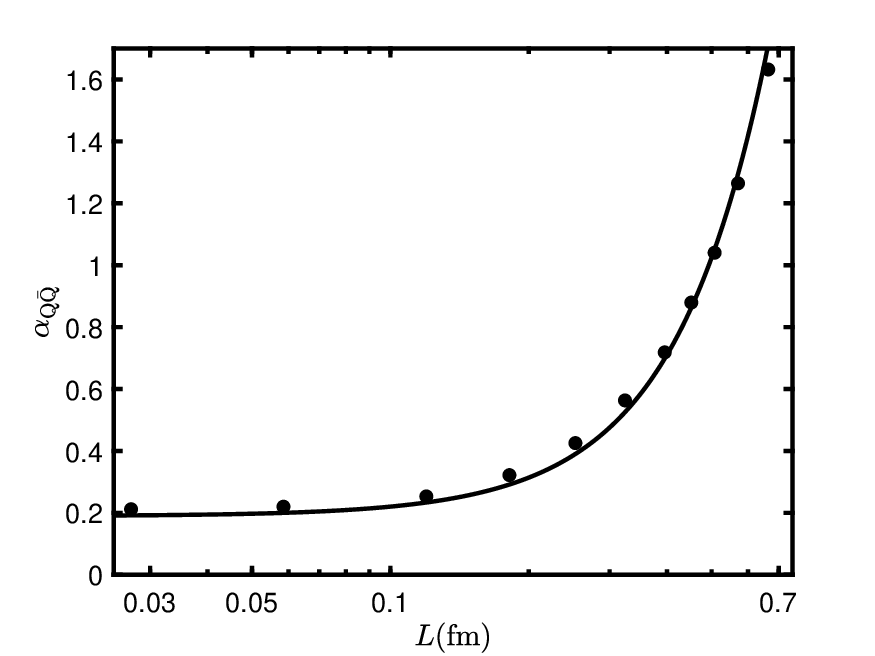}
    \caption{\label{fig2} The effective running coupling of $\mathrm{Q\Bar{Q}}$ at $T=0, \eta =0$. The black solid dot represents the lattice data from quenched QCD \cite{Kaczmarek:2004gv}.}
\end{figure}

We determine the values $\mathbf{s}=0.41 \mathrm{GeV^2}$ and $g_{\mathrm{Q\Bar{Q}}} = 0.176$ by fitting to lattice data. And the boundary condition of $r(x)$ is
\begin{equation}
    r\Big(\pm\,\frac{L}{2}\Big)=0,\,r(0)=r_{0},\,(\partial_{x}r|_{r=r_{0}})^2=0.
\end{equation}
By substituting into the Euler-Lagrange equation, we can obtain:
\begin{equation}
    \partial_{x}r=\sqrt{\frac{w(r)^{2}g_{1}(r)^{2}-g_{1}(r)w(r_{0})^{2}g_{1}(r_{0})}{w(r_{0})^{2}g_{1}(r_{0})g_{2}(r)}}.
\end{equation}
The distance between heavy quark and anti-quark is:
\begin{equation}
    L=2\int_{0}^{r_{0}}\partial_{r}x  dr,
\end{equation}
where $\partial_{r}x=\frac{\partial x}{\partial r}=\frac{1}{\partial_{x}r}$.
By $E=S/t$ and after normalizing, we can obtain the potential of $\mathrm{Q\Bar{Q}}$,
\begin{gather}
    E_{\mathrm{Q\Bar{Q}}}=2g_{\mathrm{Q\Bar{Q}}}\Big(\int_{0}^{r_{0}} \big(w(r)\sqrt{g_{1}(r)(\partial_{r}x)^2+g_{2}(r)}-\frac{1}{r^{2}}\big)dr-\frac{1}{r_{0}}\Big).
\end{gather}
We present a comparison between the potential obtained by the model and the lattice potential in Fig.~\ref{fig1}.
Therefore, the effective running coupling of $\mathrm{Q\Bar{Q}}$ from lattice QCD is given by  \cite{Kaczmarek:2004gv, Kaczmarek:2005ui,Kaczmarek:2007pb,Bazavov:2018wmo}
\begin{equation}\label{c2}
    \alpha_{\mathrm{Q\Bar{Q}}}=\frac{3L^2}{4}\frac{dE_{\mathrm{Q\Bar{Q}}}}{dL}.
\end{equation}
Then we present its basic behavior in Fig.~\ref{fig2}. When $L < 0.2,\mathrm{fm}$, the effective running coupling of $\mathrm{Q\Bar{Q}}$ tends to approach a small constant value rather than becoming arbitrarily small, with this constant corresponding to a conformal limit. At larger scales, the obvious increase of the effective coupling constant with distance also corresponds to our understanding of the strong interaction force. We discuss in detail its effective running coupling properties in moving thermal media in the subsequent sections.

\subsection{Doubly heavy baryon}\label{B}

\begin{figure}
    \centering
    \includegraphics[width=18cm]{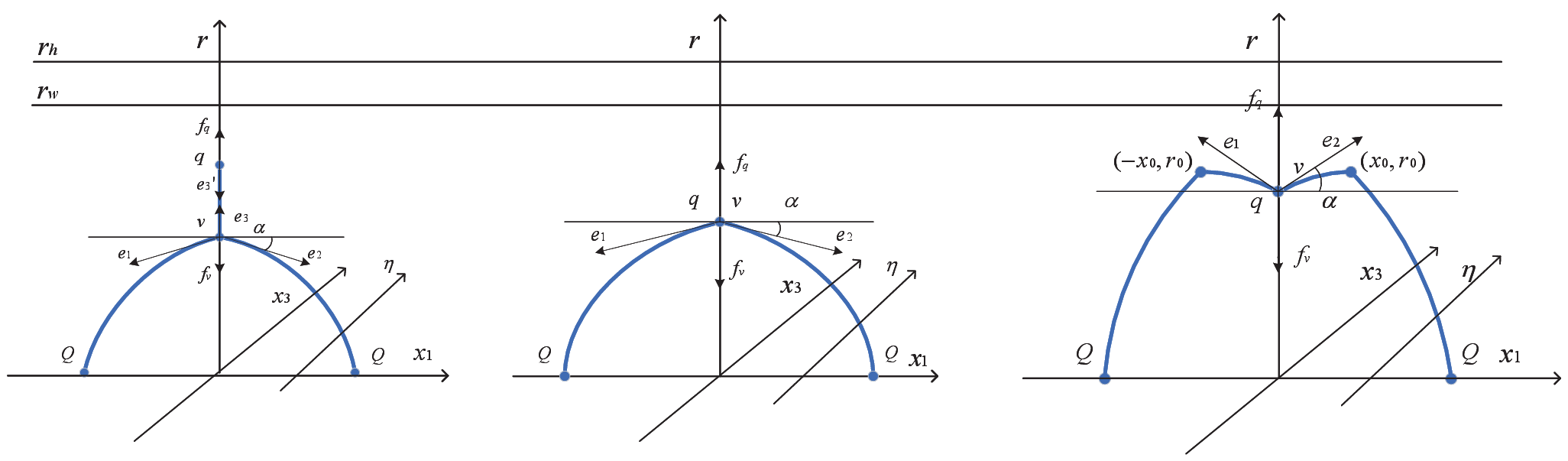}
    \caption{\label{fig3} String structure of the double heavy baryon (distance $L$ increases from left to right). The heavy quark pair is located in the $x_1$ direction, while the baryon vertex $\mathrm{v}$ and the light quark $\mathrm{q}$ are on the $r$-axis. The rapidity $\eta $ is along the $x_3$-axis direction. The heavy quark, light quark, and baryon vertex are connected by blue strings. The black arrows represent forces. $r_{h}$ is the position of the black hole horizon. $r_{w}$ is the position of an imaginary wall when the $\mathrm{QQq}$ is confined. }
\end{figure}

\begin{figure}
    \centering
    \begin{minipage}[b]{0.45\textwidth}
        \includegraphics[width=7cm]{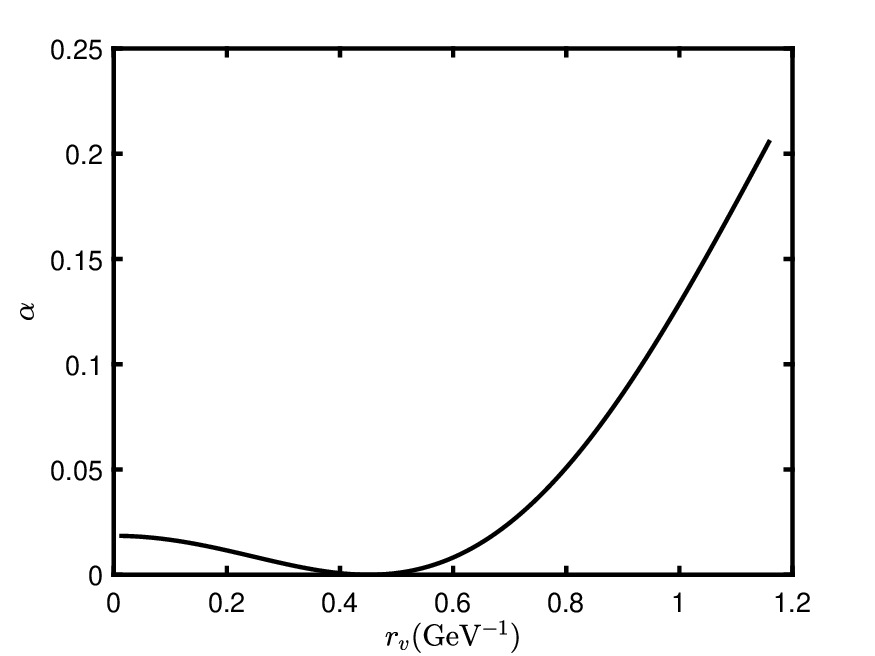}
    \end{minipage}
\hspace{0cm}
    \begin{minipage}[b]{0.45\textwidth}
        \includegraphics[width=7cm]{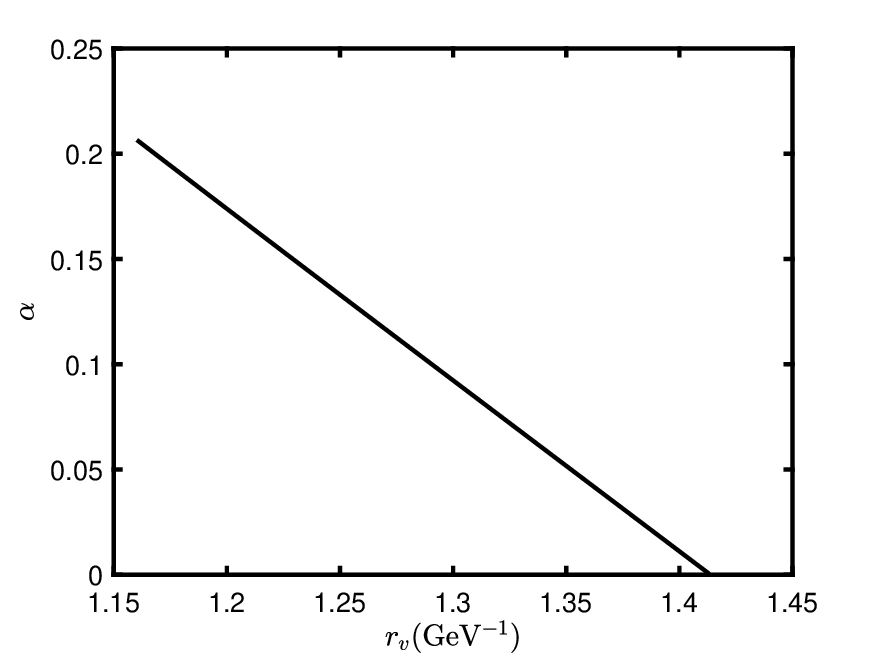}
    \end{minipage}
    \\
    \begin{minipage}[b]{0.45\textwidth}
        \includegraphics[width=7cm]{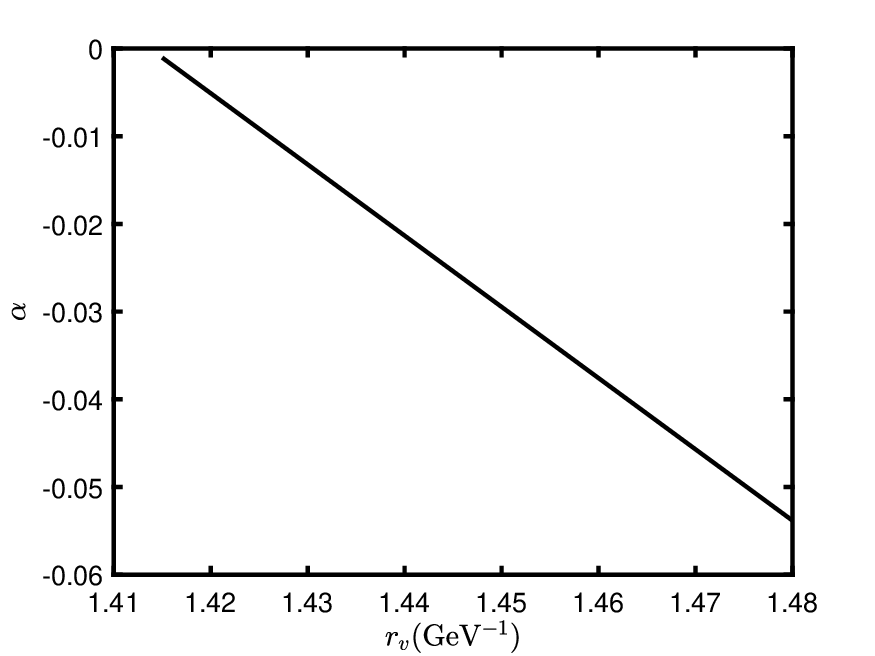}
    \end{minipage}
\hspace{0cm}
    \begin{minipage}[b]{0.45\textwidth}
        \includegraphics[width=7cm]{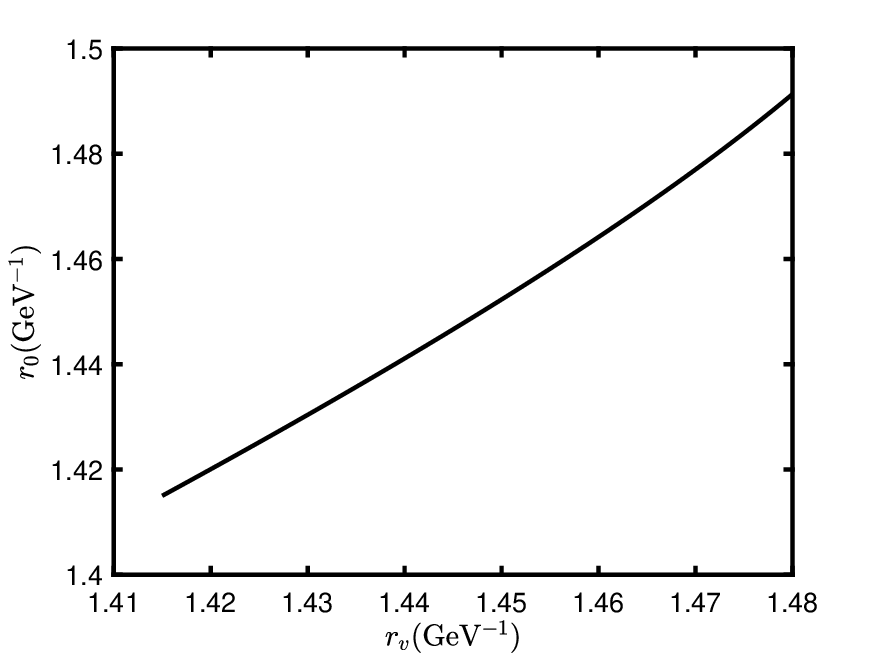}
    \end{minipage}
    \caption{\label{QQq} The top-left plot shows the string angle $\alpha$ at the baryon vertex as a function of $r_v$ for the small-$L$ configuration; the top-right plot shows $\alpha(r_v)$ for the intermediate-$L$ configuration; the bottom-left plot shows $\alpha(r_v)$ for the large-$L$ configuration; and the bottom-right plot shows $r_0$ as a function of $r_v$ for the large-$L$ configuration.}
\end{figure}

The double heavy baryon consists of two heavy quarks and one light quark, which introduces a baryon vertex. When considering a string configuration for the ground state of $\mathrm{QQq}$, it is natural due to symmetry to place the light quark between the two heavy quarks \cite{Andreev:2020xor}. The $\mathrm{QQq}$ potential depends only on the distance $L$ between the heavy quarks. In the Ref. \cite{Yamamoto:2008jz}, the $\mathrm{QQq}$ potential is also fitted to a function similar to the Cornell potential:
\begin{equation}\label{c}
    E_{\mathrm{QQq}}=\sigma_{eff}L-\frac{A_{eff}}{L}+C_{eff},
\end{equation}
where the first term represents the confinement potential, and the second term comes from one-gluon exchange \cite{Yamamoto:2008jz}. Due to the octet nature of gluons, the factor $4/3$ is included in $A_{eff}$.
As the distance between the heavy quark pair increases, its string configuration will change, as shown in Fig. \ref{fig3}. And the three configurations from left to right are called: small $L$, intermediate $L$, and large $L$. In the small $L$ configuration, the action is composed of three strings, the baryon vertex, and the light quark. After $r_v=r_q$, it transitions to the intermediate $L$ configuration, where the position of the light quark coincides with the baryon vertex, and the action is constituted by two strings, the baryon vertex, and the light quark. When the string at the baryon vertex transitions from a "convex" to a "concave" shape, it becomes the large $L$ configuration. At this configuration, the composition of the action is consistent with the intermediate $L$. Therefore, the total action at small $L$ is:
\begin{equation}
    S=\sum_{i=1}^{3} S_{NG}^{(i)}+S_{vert}+S_{q}.
\end{equation}
Then we choose the static gauge where $\xi^{0}=t,\,\xi^{1}=r$, and the boundary conditions of $x(r)$ is:
\begin{equation}
    x(0)=\pm\,\frac{L}{2},\,x(r_{v})=x(r_{q})=0,\,
    \begin{cases}
    (\partial_{r}x)^2=\cot^2(\alpha),&r=r_{v}\\
      (\partial_{r}x)^2=0,&r\in (r_{v},r_{q}].
    \end{cases}
\end{equation}
And we obtain the total action as
\begin{align}
    S=g_{\mathrm{QQq}}t&\Big(2\int_{0}^{r_{v}} w(r)\sqrt{g_{1}(r)(\partial_{r}x)^2+g_{2}(r)} dr+ \int_{r_{v}}^{r_{q}} w(r)\sqrt{g_{2}(r)}dr\notag\\
    &+3k\frac{e^{-2\mathbf{s}r^2}}{r} \sqrt{g_{1}(r)}+n\frac{e^{\frac{\mathbf{s}r^2}{2}}}{r} \sqrt{g_{1}(r)}\Big),
\end{align}
where $k=\frac{\tau_{v}}{3g_{\mathrm{QQq}}}, n=\frac{m}{g_{\mathrm{QQq}}}$. By substituting the first term of the action into the Euler-Lagrange equation, we obtain
\begin{equation}
    \mathcal{H}=\frac{w(r)g_{1}(r)\partial_{r}x}{\sqrt{g_{1}(r)(\partial_{r}x)^2+g_{2}(r)}},
\end{equation}
since $\mathcal{H}$ is a constant, we have:
\begin{equation}
    \mathcal{H}|_{r=r_{v}}=\frac{w(r_{v})g_{1}(r_{v})\cot (\alpha)}{\sqrt{g_{1}(r_{v})\cot^2(\alpha)+g_{2}(r_{v})}}.
\end{equation}
According to $\mathcal{H}=\mathcal{H}|_{r=r_{v}}$,
\begin{equation}
    \partial_{r}x=\sqrt{\frac{w(r_{v})^{2}g_{1}(r_{v})^{2}g_{2}(r)}{w(r)^{2}g_{1}(r)^{2}\big(g_{1}(r_{v})+g_{2}(r_{v})\tan^2(\alpha)\big)-g_{1}(r)w(r_{v})^{2}g_{1}(r_{v})^{2}}}.
\end{equation}
Furthermore, there must be a balance of forces at the light quark and the baryon vertex, with the light quark site satisfying $f_{q}+e_{3}'=0$, and the vertex site satisfying $f_{v}+e_{3}+e_{1}+e_{2}=0$. These forces are obtained by taking the variation of their action:
\begin{align}
    e_{1}&=g_{\mathrm{QQq}}w(r_{v})\Big(-\sqrt{\frac{g_{1}(r_{v})f(r_{v})}{f(r_{v})+\tan^{2}\alpha}},\,-\sqrt{\frac{g_{1}(r_{v})}{f(r_{v})^2\cot^2(\alpha)+f(r_{v})}}\Big),\\
    e_{2}&=g_{\mathrm{QQq}}w(r_{v})\Big(\sqrt{\frac{g_{1}(r_{v})f(r_{v})}{f(r_{v})+\tan^{2}\alpha}},\,-\sqrt{\frac{g_{1}(r_{v})}{f(r_{v})^2\cot^2(\alpha)+f(r_{v})}}\Big),\\
    e_{3}&=g_{\mathrm{QQq}}w(r_{v})\big(0,\,\sqrt{g_{2}(r_{v})}\big),\label{21}\\
    e_{3}'&=g_{\mathrm{QQq}}w(r_{q})\big(0,\,-\sqrt{g_{2}(r_{q})}\big),\\
    f_{q}&=\Big(0,\,-g_{\mathrm{QQq}}n\partial_{r_{q}}\big(\frac{e^{\frac{\mathbf{s}r_{q}^2}{2}}}{r_{q}} \sqrt{g_{1}(r_{q})}\big)\Big),\\
    f_{v}&=\Big(0,\,-3g_{\mathrm{QQq}}k\partial_{r_{v}}\big(\frac{e^{-2\mathbf{s}r_{v}^2}}{r_{v}} \sqrt{g_{1}(r_{v})}\big)\Big),\label{24}
\end{align}
it is evident that when $T$ and $\eta$ are fixed, the force at the light quark site depends only on $r_q$, while the force at the vertex involves only two unknowns, $r_v$ and $\alpha$. Therefore, we can determine the value of $r_q$ and the function $\alpha(r_v)$, where the $\alpha(r_v)$ is shown in Fig. \ref{QQq}, and from these, we can derive the potential of small $L$. The potential, with the divergent terms eliminated through normalization, is represented as:
\begin{align}
    E_{small}=g_{\mathrm{QQq}}&\Big(2\int_{0}^{r_{v}} \big(w(r)\sqrt{g_{1}(r)(\partial_{r}x)^2+g_{2}(r)}-\frac{1}{r^{2}}\big)dr-\frac{2}{r_{v}}+\int_{r_{v}}^{r_{q}} w(r)\sqrt{g_{2}(r)}dr\notag\\
    &+3k\frac{e^{-2\mathbf{s}r_{v}^2}}{r_{v}} \sqrt{g_{1}(r_{v})}+n\frac{e^{\frac{\mathbf{s}r_{q}^2}{2}}}{r_{q}} \sqrt{g_{1}(r_{q})}\Big)+c.
\end{align}

At intermediate $L$, the total action is given by
\begin{equation}
    S=\sum_{i=1}^{2} S_{NG}^{(i)}+S_{vert}+S_{q},
\end{equation}
and the boundary conditions of $x(r)$ become
\begin{equation}
    x(0)=\pm\,\frac{L}{2},\,x(r_{v})=0,\,(\partial_{r}x|_{r=r_{v}})^2=\cot^2(\alpha).
\end{equation}
And only the forces at the vertices need to be considered: $f_{v}+f_{q}+e_{1}+e_{2}=0$, Since $r_q = r_v$, we only need to replace $r_q$ with $r_v$ in the force. Then we can get the $\alpha(r_v)$ at intermediate $L$, as shown in Fig. \ref{QQq}. Similarly, we can obtain its potential at intermediate $L$ as:
\begin{align}
    E_{intermediate}=g_{\mathrm{QQq}}&\Big(2\int_{0}^{r_{v}} \big(w(r)\sqrt{g_{1}(r)(\partial_{r}x)^2+g_{2}(r)}-\frac{1}{r^{2}}\big)dr-\frac{2}{r_{v}}\notag\\
    &+3k\frac{e^{-2\mathbf{s}r_{v}^2}}{r_{v}} \sqrt{g_{1}(r_{v})}+n\frac{e^{\frac{\mathbf{s}r_{v}^2}{2}}}{r_{v}} \sqrt{g_{1}(r_{v})}\Big)+c.
\end{align}
The distance between heavy quark pairs for small $L$ and intermediate $L$ is calculated using the following function
\begin{equation}
    L=2\int_{0}^{r_{v}}\partial_{r}x  dr.
\end{equation}

For fixed temperature $T$ and rapidity $\eta$, we similarly obtain the function $\alpha(r_v)$ from the $f_{v}+f_{q}+e_{1}+e_{2}=0$. A sign change in $\alpha$ indicates that the string configuration at the baryon vertex transitions from "convex" to "concave", as shown in the rightmost plot of Fig. \ref{fig3} corresponding to the transformation from top-right to bottom-left plot in Fig. \ref{QQq}. In this case, $r_v$ is no longer the maximum value in the fifth dimension, and two smooth turning points $(x_0,r_0)$ will emerge. For the string configuration at large $L$, we choose a new metric $\xi^{0}=t,\,\xi^{1}=x$, and now the boundary condition of $r(x)$ is
\begin{equation}
    r\Big(\pm\,\frac{L}{2}\Big)=0,\,r(0)=r_{v},\,\,
    \begin{cases}
    (\partial_{x}r)^2=\tan^2(\alpha),&r=r_{v}\\
      (\partial_{x}r)^2=0,&r=r_{0}.
    \end{cases}
\end{equation}
Then we need to find the function between $r_{0}$ and $r_{v}$, we can employ the Euler-Lagrange equation and incorporate the action of the string to derive the first integral.
\begin{equation}
    \mathcal{H}=\frac{w(r)g_{1}(r)}{\sqrt{g_{1}(r)+g_{2}(r)(\partial_{x}r)^2}},
\end{equation}
$\mathcal{H}$ is a constant. We bring the boundary condition to the first integral.
\begin{align}
    \mathcal{H}|_{r=r_{0}}&=w(r_{0})\sqrt{g_{1}(r_{0})},\\
    \mathcal{H}|_{r=r_{v}}&=\frac{w(r_{v})g_{1}(r_{v})}{\sqrt{g_{1}(r_{v})+g_{2}(r_{v})\tan^2(\alpha)}},\\
    w(r_{0})\sqrt{g_{1}(r_{0})}&=\frac{w(r_{v})g_{1}(r_{v})}{\sqrt{g_{1}(r_{v})+g_{2}(r_{v})\tan^2(\alpha)}}\label{39}.
\end{align}
We can get the function $r_0(r_v)$ by Eq. (\ref{39}), as shown in the bottom-right plot of Fig. \ref{QQq}. Furthermore we obtain
\begin{gather}
    \partial_{x}r=\sqrt{\frac{w(r)^{2}g_{1}(r)^{2}-g_{1}(r)w(r_{0})^{2}g_{1}(r_{0})}{w(r_{0})^{2}g_{1}(r_{0})g_{2}(r)}}.
\end{gather}
For convenience in the calculation, we present the potential after equivalent transformation and normalization as follows
\begin{align}
    E_{large}=g_{\mathrm{QQq}}&\Big(2\int_{0}^{r_{0}} \big(w(r)\sqrt{g_{1}(r)(\partial_{r}x)^2+g_{2}(r)}-\frac{1}{r^{2}}\big)dr+2\int_{r_{v}}^{r_{0}} w(r)\sqrt{g_{1}(r)(\partial_{r}x)^2+g_{2}(r)}dr\notag\\
    &-\frac{2}{r_{v}}
    +3k\frac{e^{-2\mathbf{s}r_{v}^2}}{r} \sqrt{g_{1}(r_{v})}+n\frac{e^{\frac{\mathbf{s}r_{v}^2}{2}}}{r_{v}} \sqrt{g_{1}(r_{v})}\Big)+c.
\end{align}
The distance at large $L$ should be
\begin{equation}
    L=2\Big(\int_{0}^{r_{0}}\partial_{r}x dr+\int_{r_{v}}^{r_{0}}\partial_{r}x dr\Big).
\end{equation}
Clearly, the complete potential is pieced together from three potential functions, but we do not need to focus on which configuration it belongs to. Therefore, in the following text, the potential of $\mathrm{QQq}$ will be collectively referred to as $E_{\mathrm{QQq}}$. For detailed computational procedures of the QQq potential, refer to Refs. \cite{Liu:2023hoq, Andreev:2020xor}.

\begin{figure}
    \centering
    \includegraphics[width=8.5cm]{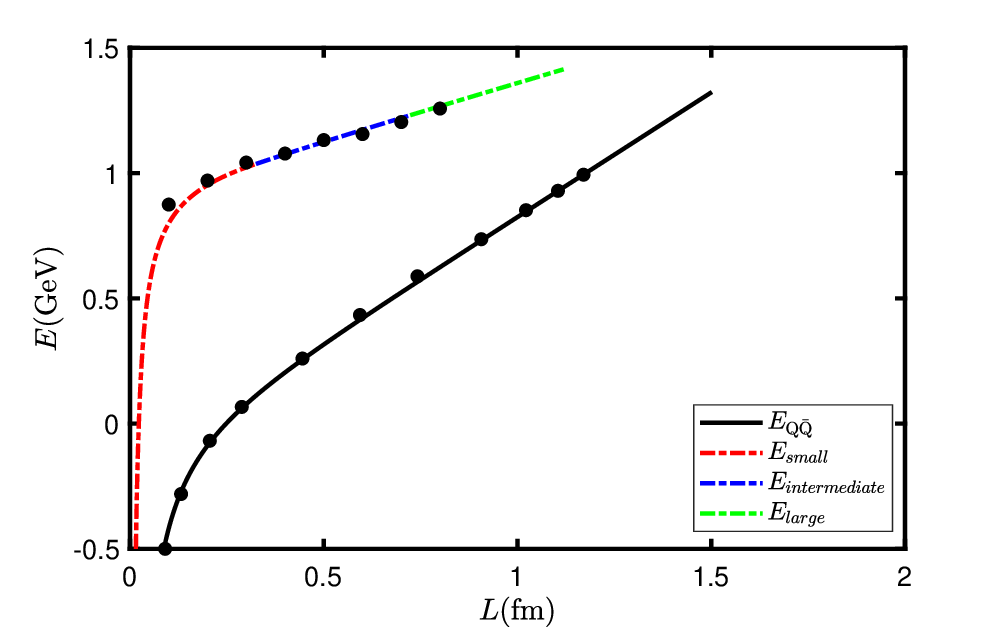}
    \caption{\label{fig4} The potential of $\mathrm{QQq}$ and $\mathrm{Q\Bar{Q}}$ at $T=0, \eta =0$. The solid black lines represent the $\mathrm{Q\Bar{Q}}$ potential. The curves composed of dotted lines in three colors represent the $\mathrm{QQq}$ potential, with red indicating small $L$, blue for intermediate $L$, and green representing large $L$. The black solid dot represents the lattice data from quenched QCD\cite{Kaczmarek:2004gv, Yamamoto:2008jz, Najjar:2009da}.}
\end{figure}

The parameters determined through fitting the lattice points for the $\mathrm{QQq}$ potential are: $\mathbf{s}=0.41 \mathrm{GeV^2}, k=-0.321, n=2.941, g_{\mathrm{QQq}}=0.082, c=0.73 \mathrm{GeV}$. The potential of $\mathrm{QQq}$ is shown in Fig. \ref{fig4}. The last point for small $L$ is $(r_v, L, E) = (1.1600, 0.3250, 1.0365)$, the starting point for intermediate $L$ is $(r_v, L, E) = (1.1602, 0.3251, 1.0366)$, and its last point is $(r_v, L, E) = (1.4132, 0.7174, 1.2282)$. The starting point for large $L$ is $(r_v, L, E) = (1.4150, 0.7233, 1.2310)$, with no sudden change in the data.

\begin{figure}
    \centering
    \includegraphics[width=8.5cm]{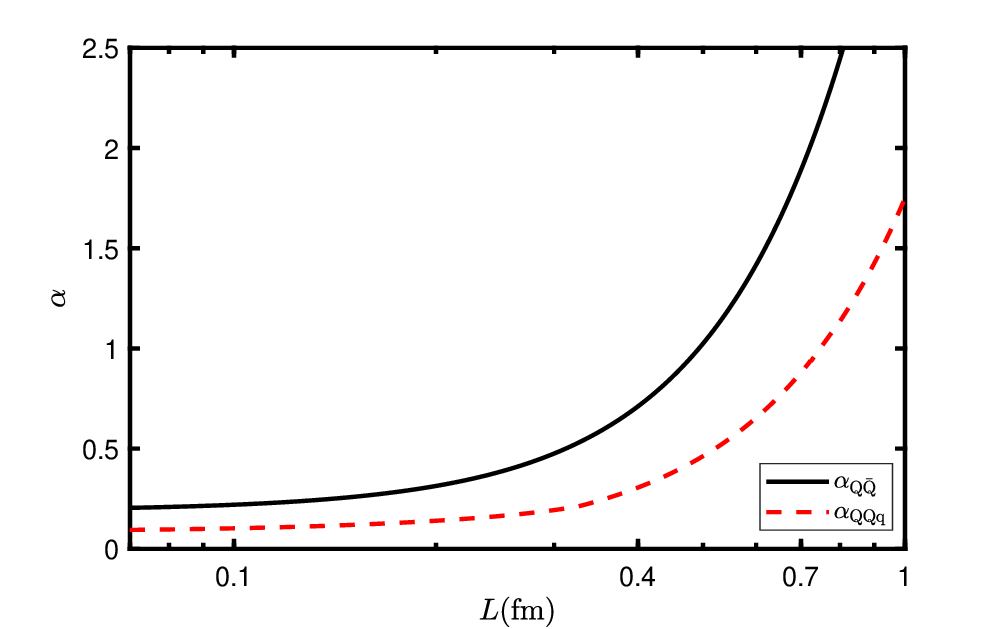}
    \caption{\label{fig5} The effective running coupling of $\mathrm{QQq}$ and $\mathrm{Q\Bar{Q}}$ at $T=0, \eta =0$. The red dashed line represents the $\mathrm{QQq}$, while the black solid line represents the $\mathrm{Q\Bar{Q}}$.}
\end{figure}

Based on the model of heavy quarkonium in Refs. \cite{Kaczmarek:2004gv, Kaczmarek:2005ui,Kaczmarek:2007pb,Bazavov:2018wmo}, similarly fitted to the Cornell potential for the $\mathrm{QQq}$, we can also determine the effective running coupling,
\begin{equation}\label{c1}
    \alpha_{\mathrm{QQq}} = \frac{3L^2}{4} \frac{dE_{\mathrm{QQq}}}{dL},
\end{equation}
to study the force between heavy quarks. It is the inter-two-quark potential in baryons which effectively includes the light-quark effects \cite{Yamamoto:2008jz}. Moreover, it can be proven that the physical significance of Eq. (\ref{c2}) and Eq. (\ref{c1}) is the same, as both represent the effective coupling strength between heavy quarks. Thus, the effective running coupling of $\mathrm{QQq}$ is a function of the separation distance between the two heavy quarks as shown in Fig. \ref{fig5}. It can be seen that at small scales, the interaction between quarks in the $\mathrm{QQq}$ becomes very weak and flat, and its range $(L< 0.4\,\mathrm{fm})$ is broader compared to $\mathrm{Q\Bar{Q}}$. Moreover, the effective coupling of $\mathrm{QQq}$ is always smaller than that of $\mathrm{Q\Bar{Q}}$ and is almost half of it. This is very close to the relationship between $A_{\mathrm{QQq}}$ and $A_{\mathrm{Q\Bar{Q}}}$ fitted in the lattice \cite{Yamamoto:2008jz}, which is due to the presence of the light quark reducing the interquark force \cite{Yamamoto:2008fm}.

\section{Numerical Results and Discussion}\label{sec:3}

\begin{figure}
    \centering
    \includegraphics[width=8.5cm]{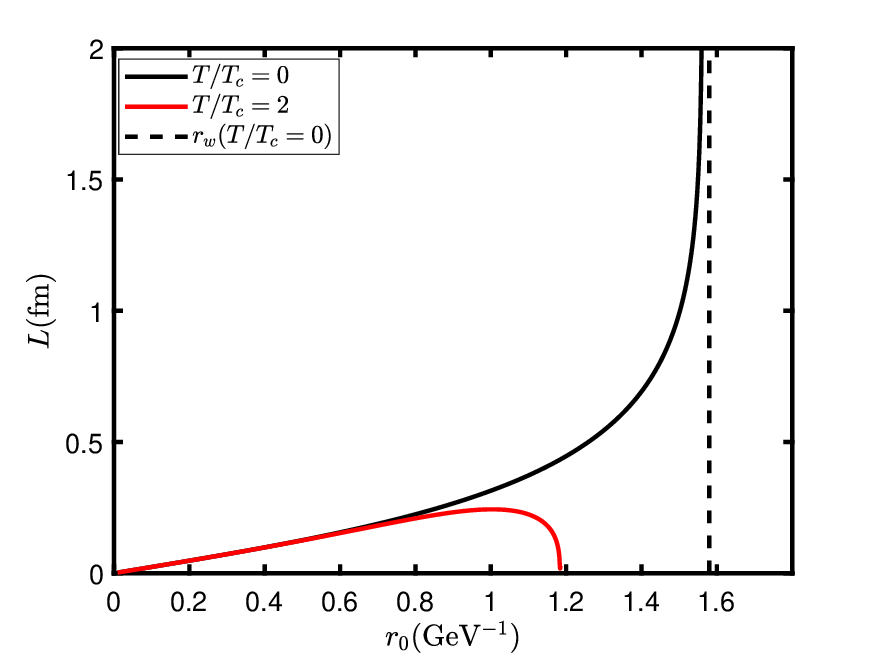}
    \caption{\label{QQrw} The black curve represents the inter-heavy-quark distance of $\mathrm{Q\Bar{Q}}$ at $T/T_c=0, \eta =0$, and the red curve represents the inter-heavy-quark distance of $\mathrm{Q\Bar{Q}}$ at $T/T_c=2, \eta =0$. The black dotted line represents the imaginary wall.}
\end{figure}

\begin{figure}
    \centering
    \includegraphics[width=10cm]{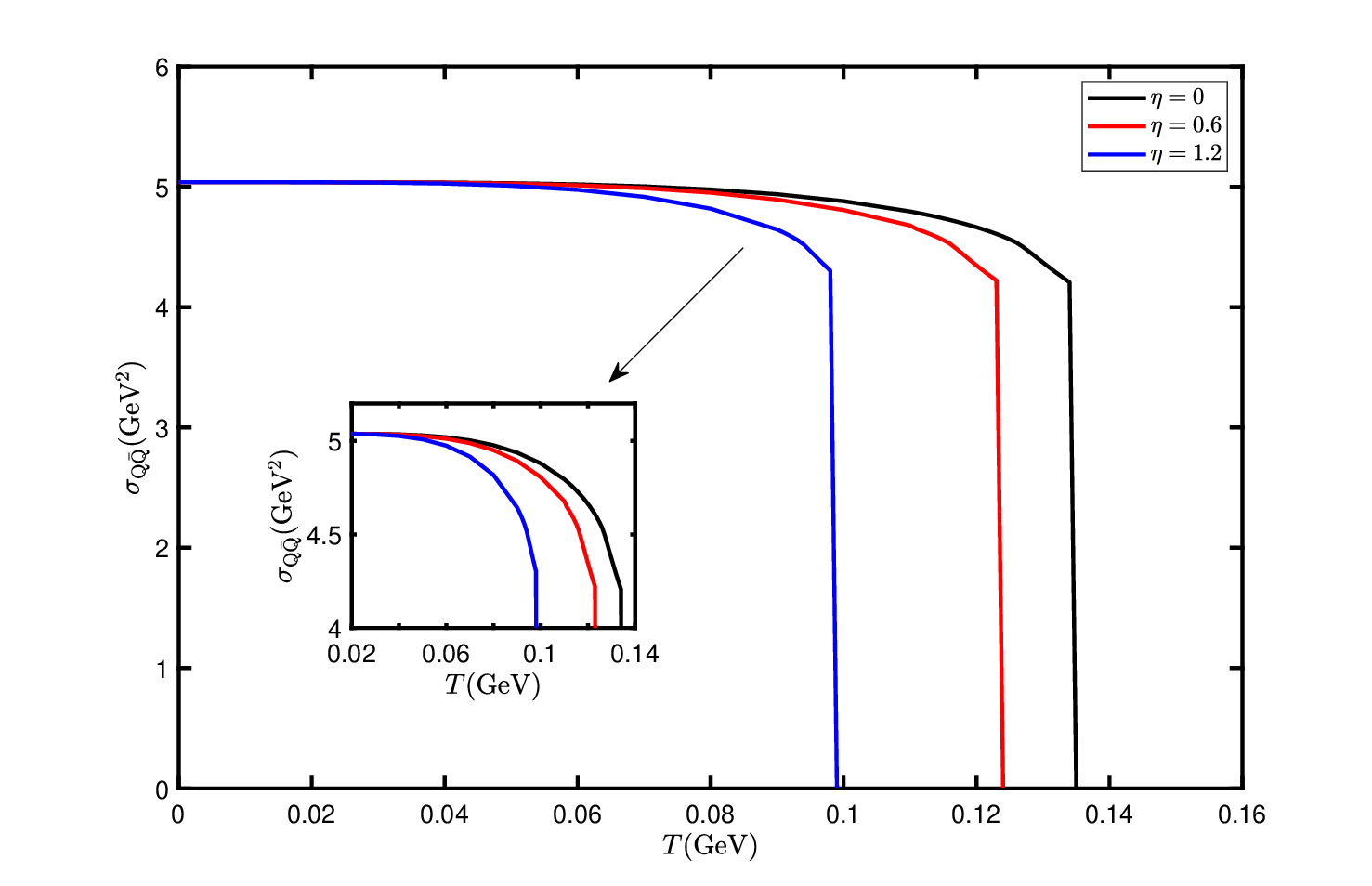}
    \caption{\label{st1} The temperature dependence of the string tension for $\mathrm{Q\Bar{Q}}$, where the black line represents $\eta =0$, the red line represents $\eta =0.6$, and the blue line represents $\eta =1.2$.}
\end{figure}

Before we proceed with the discussion, we need to obtain the critical points (critical temperatures at different rapidities). At low temperatures and rapidities, $\mathrm{Q\Bar{Q}}$ is confined, and an "imaginary wall" exists at $r = r_w$. The imaginary wall is solely determined by the background and characterizes the highest reachable point in the fifth dimension for string in the confinement phase, where the $L$ diverges. For the $\mathrm{Q\Bar{Q}}$, the imaginary wall naturally corresponds to the value of $r_0$ at which $L$ diverges, i.e., the maximum attainable value of $r_0$. For the QQq with large $L$, the maximum five-dimensional coordinate of the string is actually $r_0$, meaning the imaginary wall position $r_w$ corresponds to the maximum attainable value of $r_0$. However, since $r_0$ and $r_v$ are related and can be determined in terms of each other, we choose to define the position of the imaginary wall $r_w$ for QQq as the maximum value of $r_v$. The potential of the confined $\mathrm{Q\Bar{Q}}$ increases with distance, but when it rises to a certain value, string breaking occurs, exciting light quark and anti-quark from the vacuum, with quarks always remaining confined within a hadron. At high temperatures and rapidities, $\mathrm{Q\Bar{Q}}$ becomes deconfined, and the imaginary wall disappears. At this point, there is a maximum quark separation distance. When this distance is exceeded, the $\mathrm{Q\Bar{Q}}$ no longer forms a U-shaped string configuration but instead consists of two straight string segments extending from the boundary to the horizon \cite{Finazzo:2014rca}. At this point, the potential also reaches its maximum, indicating that the $\mathrm{Q\Bar{Q}}$ is screened at this point. This distance is known as the screening length. Considering string breaking or screening as a characteristic to distinguish between confinement and deconfinement, we can obtain the critical point of $\mathrm{Q\Bar{Q}}$. Furthermore, the string tension provides a significant criterion for identifying critical points, as demonstrated in Ref. \cite{Yang:2015aia}:
\begin{equation}
    \sigma _{\mathrm{Q\Bar{Q}}}=\frac{dE_{\mathrm{Q\Bar{Q}}}}{dL} |_{r=r_{w}}.
\end{equation}
Here, $r_w$ denotes the position of the imaginary wall. When the $\mathrm{Q\Bar{Q}}$ is confined, the inter-heavy-quark separation $L$ diverges as the string turning point $r_0$ approaches $r_w$, as illustrated in Fig. \ref{QQrw}. The temperature dependence of the $\sigma _{\mathrm{Q\Bar{Q}}}$ for various rapidities is shown in Fig. \ref{st1}. The string tension exhibits a monotonic decrease with increasing temperature, followed by an abrupt disappearance at the critical temperature. This is because string tension characterizes the long-range interaction between quarks in the confined phase, but when $\mathrm{Q\Bar{Q}}$ becomes deconfined, screening occurs at long distances, and quarks become free quarks.

The critical points for $\mathrm{QQq}$ can be determined similarly, however, it possesses an additional property. At high temperature and/or rapidity $\mathrm{QQq}$ can dissociate at a certain distance. However, At extreme high temperature and/or rapidity the $\mathrm{QQq}$ can not exist judged from the force balance. These maximum $(T,\eta)$ points can be obtained through Eqs. (\ref{f1}).
\begin{equation}\label{f1}
    \begin{cases}
       f_{q}+e_{3}'=0
       \\
     \frac{\partial (f_{q}+e_{3}')}{\partial r_{q}}=0
    \end{cases}.
\end{equation}

\begin{figure}
    \centering
    \includegraphics[width=8.5cm]{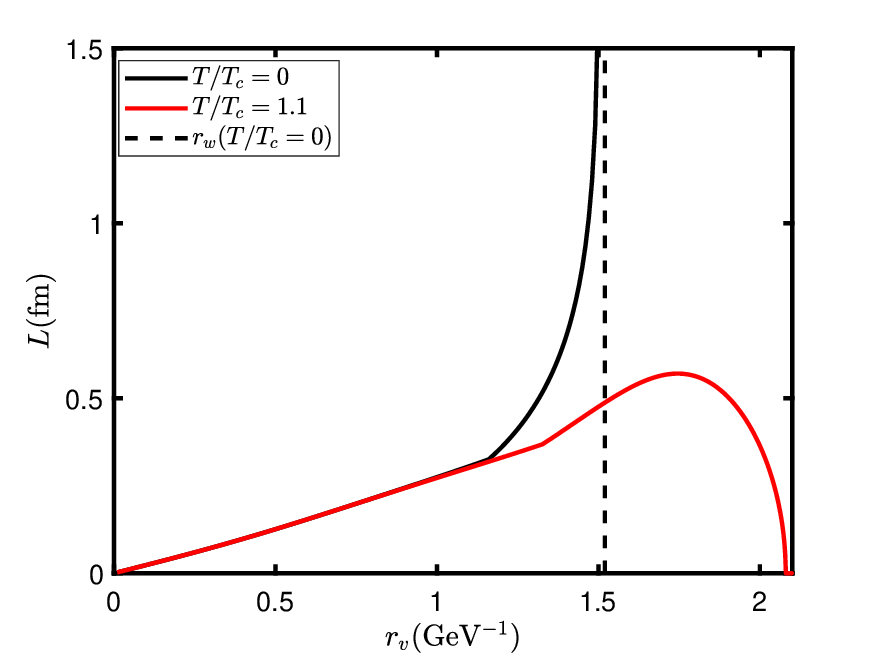}
    \caption{\label{QQqrw} The black curve represents the inter-heavy-quark distance of $\mathrm{QQq}$ at $T/T_c=0, \eta =0$, and the red curve represents the inter-heavy-quark distance of $\mathrm{QQq}$ at $T/T_c=1.1, \eta =0$. The black dotted line represents the imaginary wall.}
\end{figure}

\begin{figure}
    \centering
    \includegraphics[width=10cm]{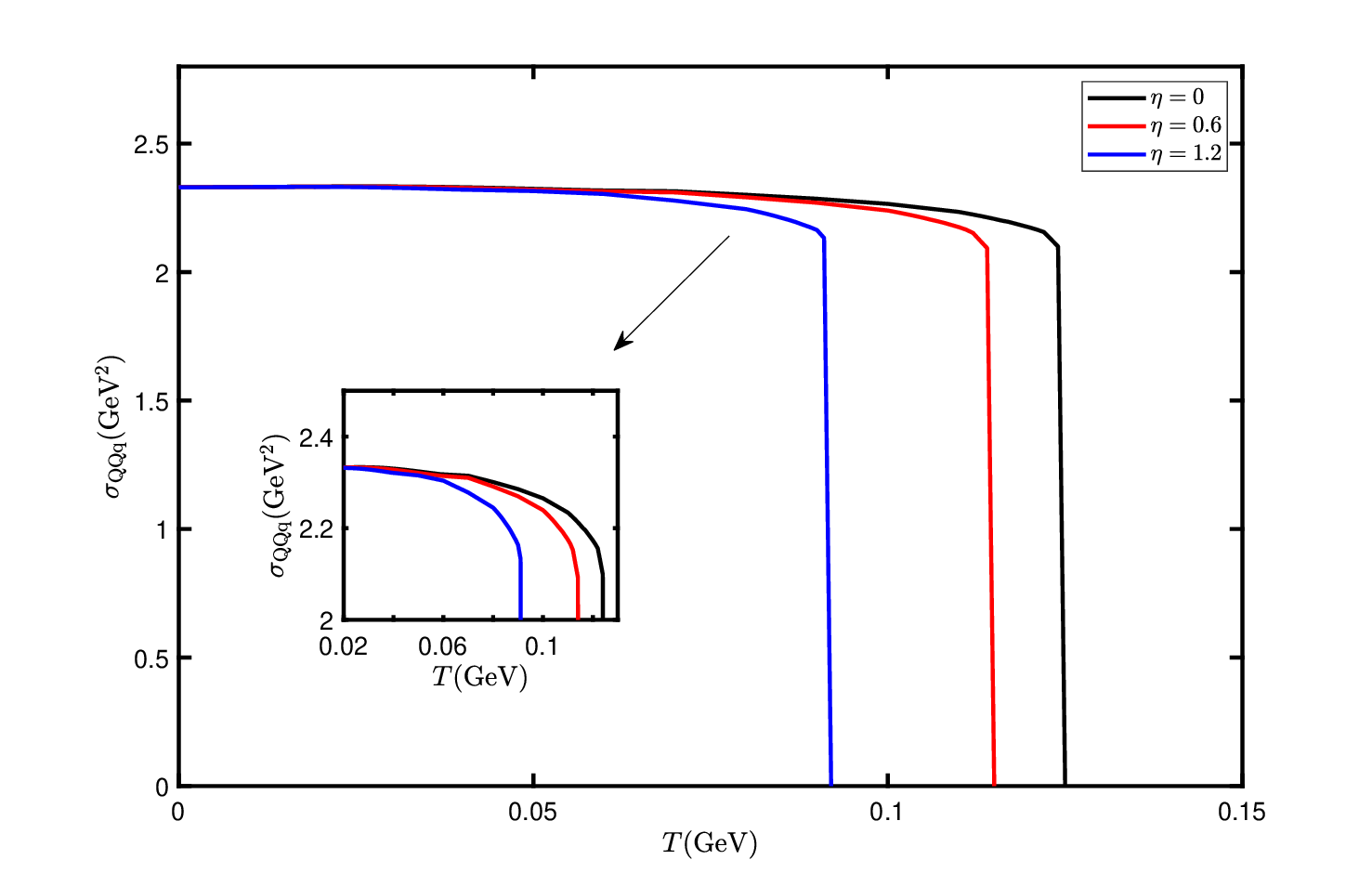}
    \caption{\label{st2} The temperature dependence of the string tension for QQq, where the black line represents $\eta =0$, the red line represents $\eta =0.6$, and the blue line represents $\eta =1.2$.}
\end{figure}

\begin{figure}
    \centering
    \includegraphics[width=8.5cm]{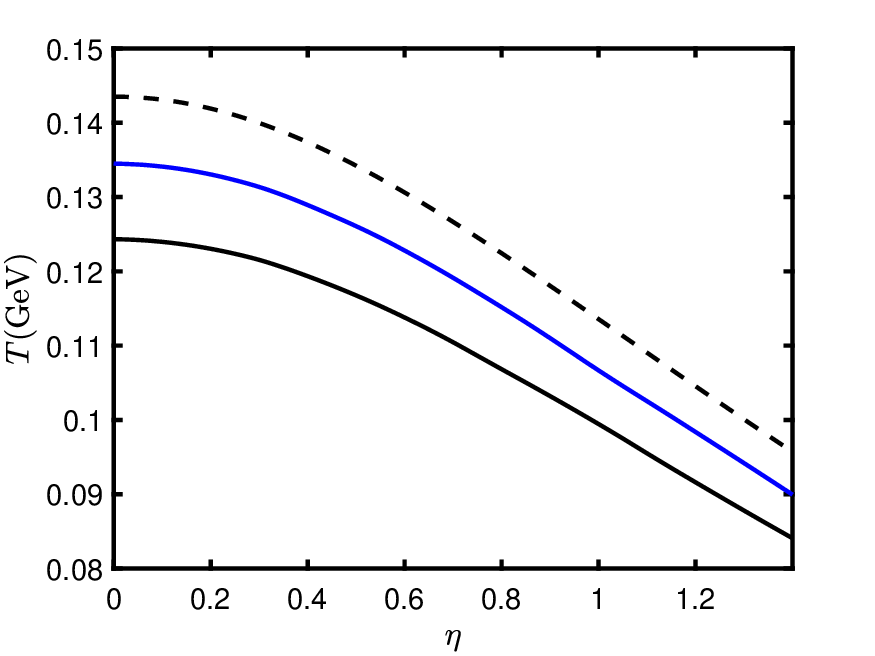}
    \caption{\label{fig6} The black solid line is composed of the critical points of $\mathrm{QQq}$, the black dashed line is composed of the maximum $(T, \eta)$ points of $\mathrm{QQq}$; the blue solid line is composed of the critical points of $\mathrm{Q\Bar{Q}}$.}
\end{figure}

From the Cornell potential perspective of QQq, $A_{eff}$ characterizes the interaction strength at short distances, while $\sigma_{eff}$ reflects the long-range interaction strength. Building upon our previous definition of the effective coupling for QQq through $A_{eff}$, we can analogously define its string tension
\begin{equation}
    \sigma _{\mathrm{QQq}}=\frac{dE_{\mathrm{QQq}}}{dL} |_{r=r_{w}}.
\end{equation}
The confined QQq also possesses an imaginary wall. Its inter-heavy-quark distance $L$ similarly diverges when the baryon vertex $r_v$ approaches $r_w$, as demonstrated in Fig. \ref{QQqrw}. And the temperature dependence of the $\sigma _{\mathrm{QQq}}$ is plotted in Fig. \ref{st2}. Similarly, the $\sigma _{\mathrm{QQq}}$ decreases with increasing temperature and drops to zero at critical temperature. And just like the effective coupling, the string tension of QQq is significantly lower than that of $\mathrm{Q\Bar{Q}}$. Base on the previous discussion, we can draw out the state diagrams of $\mathrm{Q\Bar{Q}}$ and $\mathrm{QQq}$ on the $(T, \eta)$ plane as in Fig. \ref{fig6}.

\subsection{Discussion about heavy quarkonium}

\begin{figure}[ht]
    \centering
    \begin{minipage}{0.45\textwidth}
        \includegraphics[width=7cm]{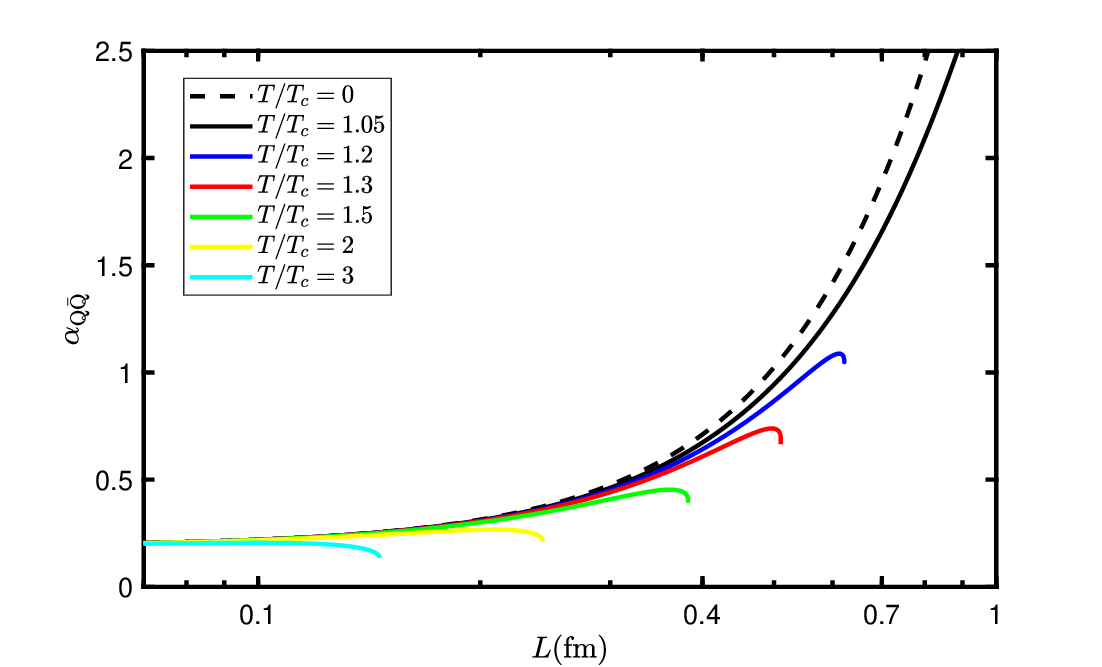}
    \end{minipage}%
    \begin{minipage}{0.45\textwidth}
        \includegraphics[width=7cm]{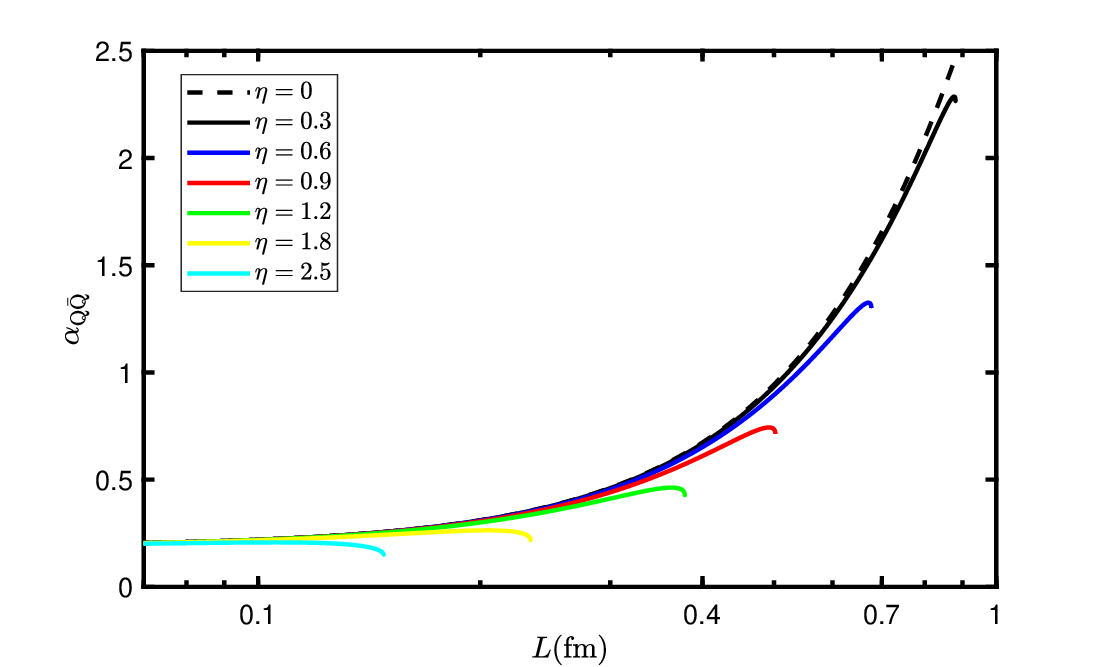}
    \end{minipage}
    \caption{\label{fig7} The left shows the temperature dependence of the effective running coupling for $\mathrm{Q\Bar{Q}}$, where $\eta=0, T_c=0.1345 \mathrm{GeV}$. The right shows the rapidity dependence of the effective running coupling for $\mathrm{Q\Bar{Q}}$, at $T=0.1412\,\mathrm{GeV}$. These plots terminate at the screening distance.}

\end{figure}

Next, we discuss the temperature dependence of the effective running coupling of $\mathrm{Q\Bar{Q}}$. From Fig. \ref{fig6}, it can be seen that critical temperature $T_c=0.1345 \mathrm{GeV}$ when $\eta=0$. We calculate the effective running coupling in the range of $T/T_c \in [0, 3]$ as shown in Fig. \ref{fig7}. When $T/T_c \in [0, 1]$, $\mathrm{Q\Bar{Q}}$ is in a confined state and the effect of temperature on the effective running coupling is relatively small. Therefore, our discussion of effective running coupling primarily concentrates on the deconfined state when $T/T_c > 1$. Additionally, the string breaking can happen when $T/T_c < 1$. The detailed discussion of string breaking at finite temperature and rapidity can be found in our previous works \cite{Liu:2023hoq}.

From the left of Fig.\ref{fig7}, it can be observed that the higher the temperature, the smaller the effective running coupling, and at small scales, the effective running coupling's dependence on temperature is minor. Additionally, the effect of temperature on effective running coupling primarily manifests in the maximum effective coupling constant $\tilde{\alpha}_{\mathrm{Q\Bar{Q}}} (T, \eta)$, with the $\tilde{\alpha}_{\mathrm{Q\Bar{Q}}}$ decreasing as the temperature increases, and the maximum coupling distance $L_{\tilde{\alpha}_{\mathrm{Q\Bar{Q}}}}$ reduces. The impact of rapidity on the effective running coupling is similar to that of temperature, as shown in the right of Fig. \ref{fig7}. We choose a lower temperature, $T=0.1412\,\mathrm{GeV}$ (when $\eta=0, T=1.05T_c$), to observe a comprehensive process of the influence of rapidity on the effective running coupling. It can be seen that the greater the rapidity, the smaller the effective running coupling, and the smaller the $\tilde{\alpha}_{\mathrm{Q\Bar{Q}}}$. Furthermore, it is noteworthy that the screening distance $L_{screen}$ for $\mathrm{Q\Bar{Q}}$ is greater than $L_{\tilde{\alpha}_{\mathrm{Q\Bar{Q}}}}$. Here we provide the definition: $\bigtriangleup L= L_{screen}-L_{\tilde{\alpha}_{\mathrm{Q\Bar{Q}}}}$.

\begin{figure}
    \centering
    \begin{minipage}{0.45\textwidth}
        \includegraphics[width=7cm]{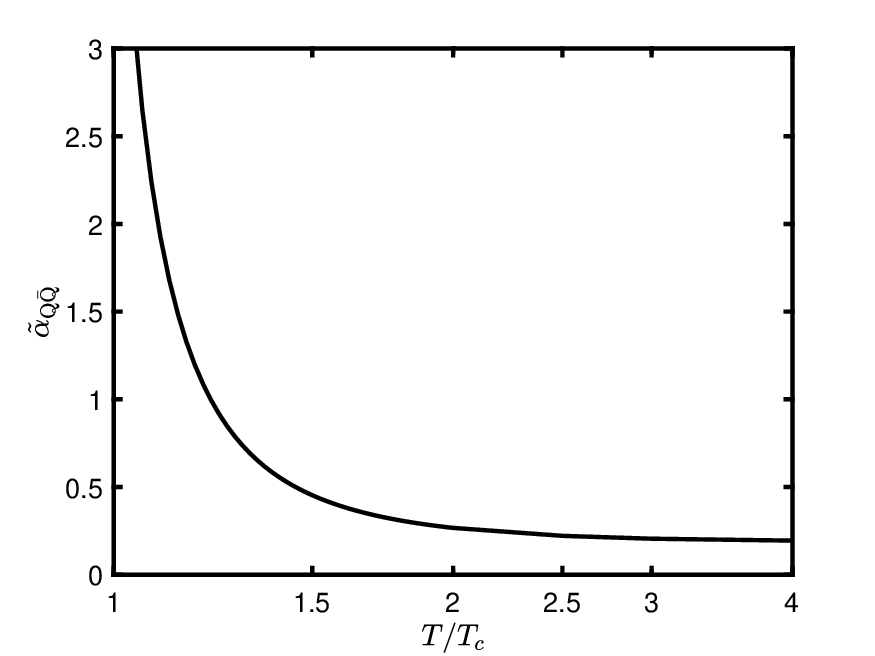}
    \end{minipage}%
    \begin{minipage}{0.45\textwidth}
        \includegraphics[width=7cm]{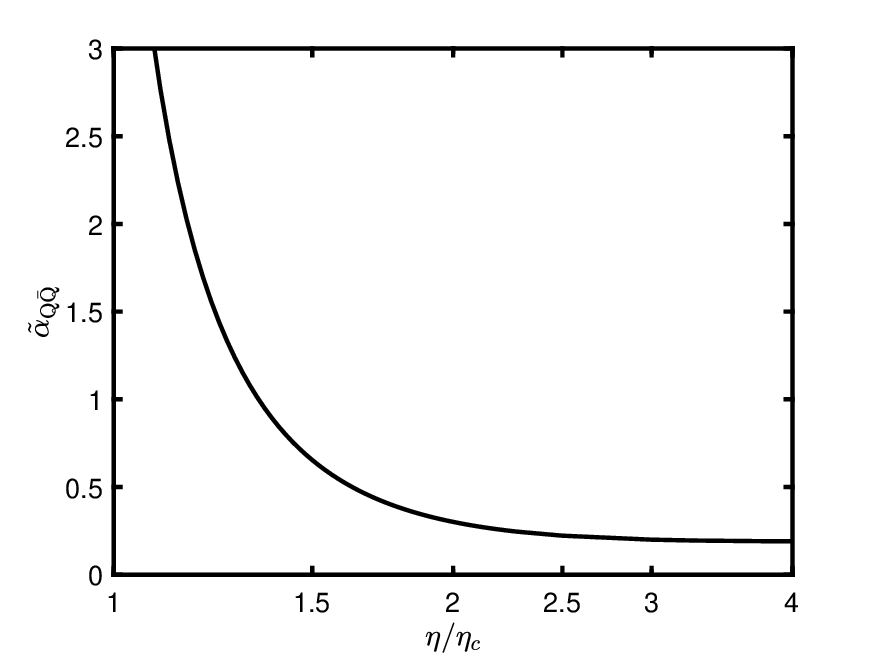}
    \end{minipage}
    \caption{\label{fig8} The left graph represents the maximum coupling $\tilde{\alpha}_{\mathrm{Q\Bar{Q}}}$ as a function of $T/T_c$ at $\eta=0$. The right graph represents the maximum coupling as a function of $\eta/\eta_c$ at $T=0.1 \mathrm{GeV}$. }
\end{figure}

\begin{figure}
    \centering
    \begin{minipage}[b]{0.45\textwidth}
        \includegraphics[width=7cm]{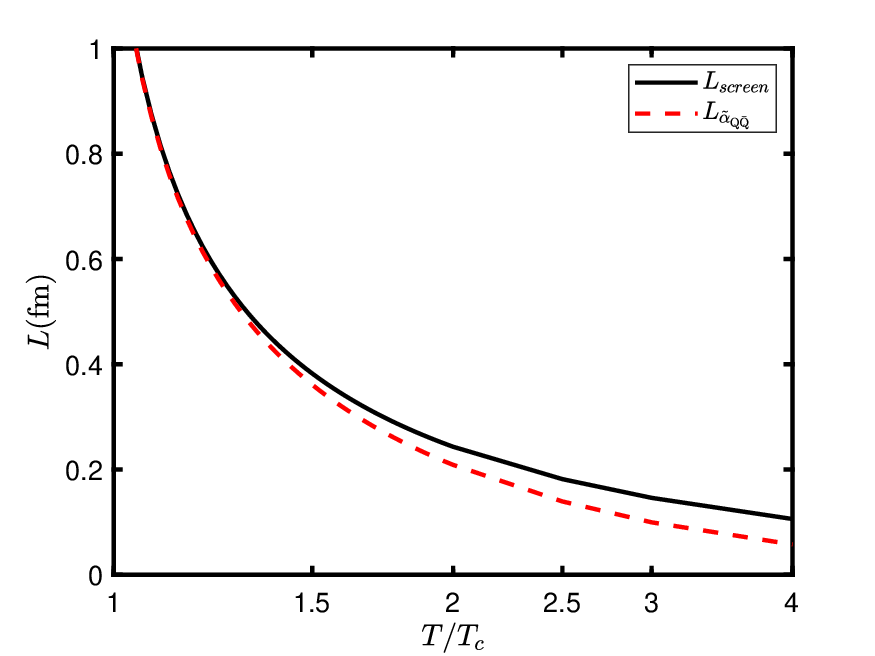}
    \end{minipage}
\hspace{0cm}
    \begin{minipage}[b]{0.45\textwidth}
        \includegraphics[width=7cm]{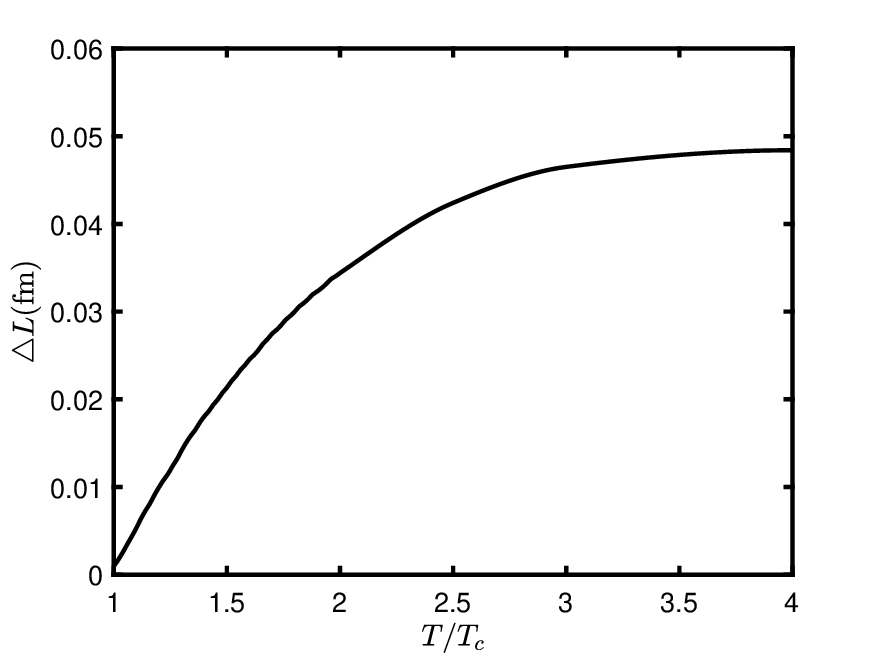}
    \end{minipage}
    \\
    \begin{minipage}[b]{0.45\textwidth}
        \includegraphics[width=7cm]{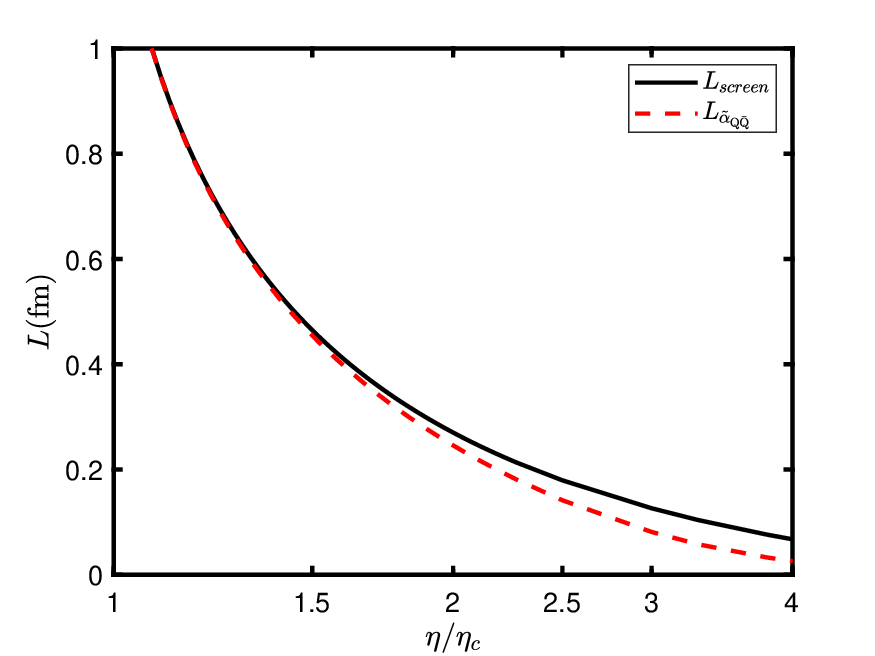}
    \end{minipage}
\hspace{0cm}
    \begin{minipage}[b]{0.45\textwidth}
        \includegraphics[width=7cm]{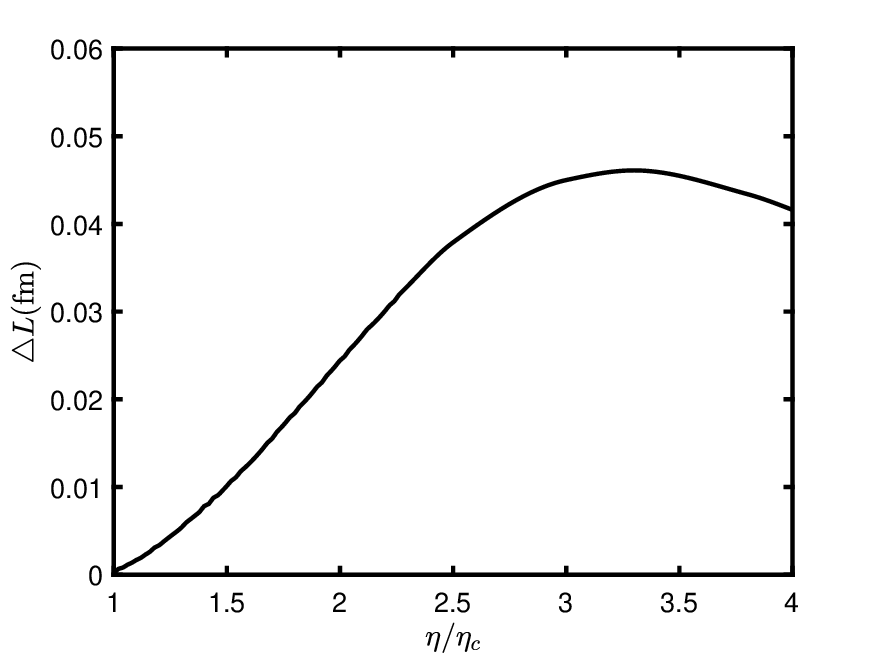}
    \end{minipage}
    \caption{\label{fig9} In the composite image, the left two images show the screening distance for $\mathrm{Q\Bar{Q}}$ (solid black lines) and the maximum coupling distance (red dotted lines). The top image is plotted against $T/T_c$ at a fixed $\eta =0$, and the bottom image against $\eta/\eta_c$ at a fixed $T=0.1 \mathrm{GeV}$. On the right, the two images illustrate the difference between these distances ($\bigtriangleup L$).}
\end{figure}

The maximum effective coupling constant as a function of $T/T_c$($\eta=0$) and $\eta/\eta_c$($T=0.1 \mathrm{GeV}$) is shown in Fig. \ref{fig8}. From the left graph, it is apparent that when $T/T_c < 1.5$, the maximum effective coupling constant rapidly decreases to a lower level as the temperature increases, and becomes relatively flat for $T/T_c > 2$. The rapidity dependence of the maximum effective coupling constant is qualitatively similar to that of temperature; however, as a function of rapidity, it exhibits a more gradual decline and a broader range $(\eta/\eta_c <2)$ in the falling region. We present the screening distance and the maximum coupling distance in Fig. \ref{fig9}. The screening distance decreases with increasing temperature or rapidity, while the slope also decreases with increasing temperature or rapidity. The screening distance as a function of $T/T_c$ behaves similarly to that observed in the Ref. \cite{Kaczmarek:2005zn}.  The $\bigtriangleup L$ increases with temperature and becomes conspicuously flat, and as a function of $\eta/\eta_c$, it displays a declining trend for $\eta/\eta_c >3.5$.

\begin{figure}
    \centering
    \includegraphics[width=8.5cm]{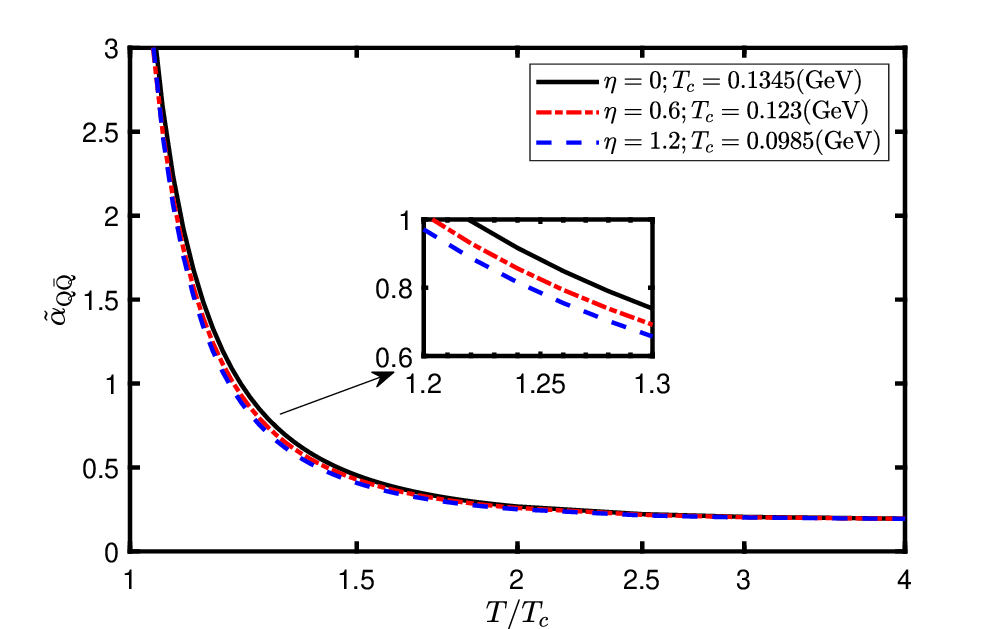}
    \caption{\label{fig10} The maximum effective coupling constant $\tilde{\alpha}_{\mathrm{Q\Bar{Q}}}$ as a function of $T/T_c$, where the solid black line represents $\eta =0$, the red dotted line represents $\eta =0.6$, and the blue dashed line represents $\eta =1.2$.}
\end{figure}

\begin{figure}[ht]
    \centering
    \begin{minipage}{0.45\textwidth}
        \includegraphics[width=7cm]{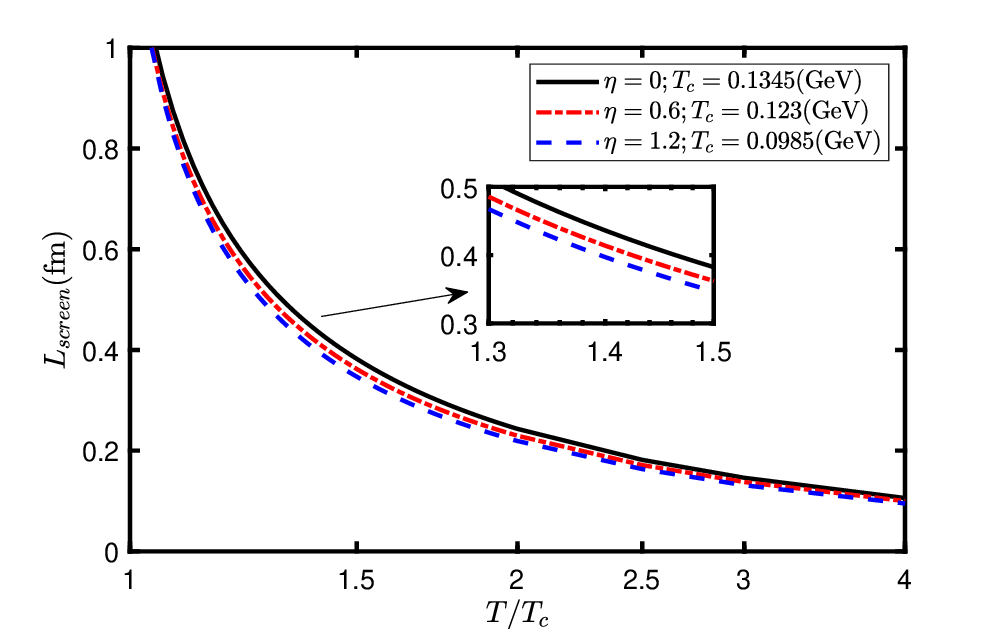}
    \end{minipage}%
    \begin{minipage}{0.45\textwidth}
        \includegraphics[width=7cm]{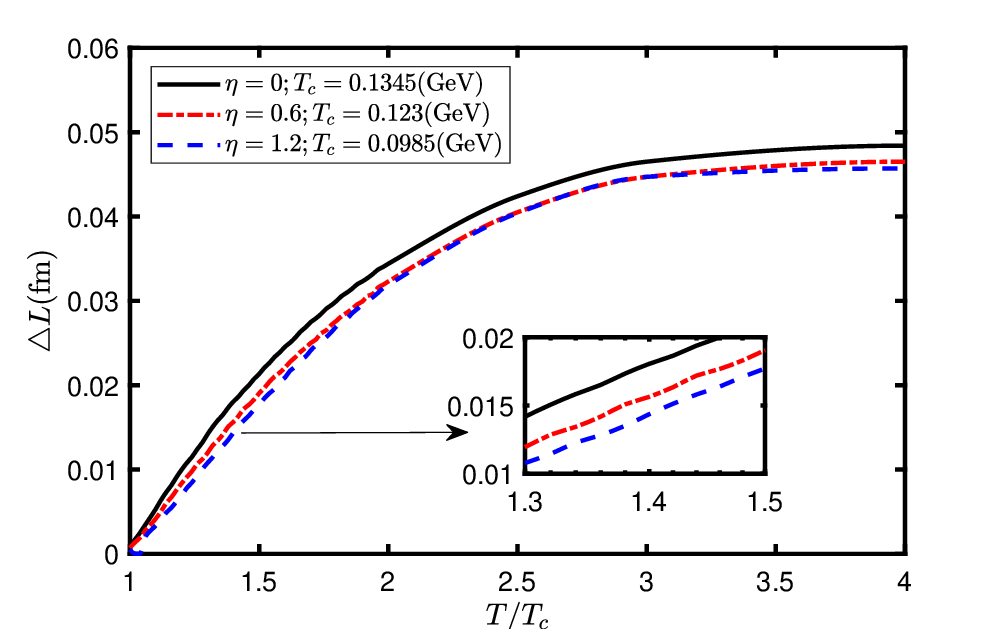}
    \end{minipage}
    \caption{\label{fig11} The left graph represents the screening distance as a function of $T/T_c$, and the right graph represents the $\bigtriangleup L$ as a function of $T/T_c$, where the solid black line represents $\eta =0$, the red dotted line represents $\eta =0.6$, and the blue dashed line represents $\eta =1.2$.}

\end{figure}

Next, we select three sets of data from the critical points in Fig. \ref{fig6} for comparison: $(\eta, T_c)=(0, 0.1345)$, $(0.6, 0.123)$, $(1.2, 0.0985)$. From Fig. \ref{fig10} and Fig. \ref{fig11}, it can be seen that although they both represent the function of the maximum effective coupling constant or screening distance with respect to $T/T_c$, there are slight differences under different rapidities. Specifically, the larger the rapidity, the smaller the maximum effective coupling constant and screening distance. This suggests that the larger the rapidity, the stronger the dependence of the maximum effective coupling constant and screening distance on temperature.

\subsection{Discussion about doubly heavy baryon}
\begin{figure}[ht]
    \centering
    \begin{minipage}{0.45\textwidth}
        \includegraphics[width=7cm]{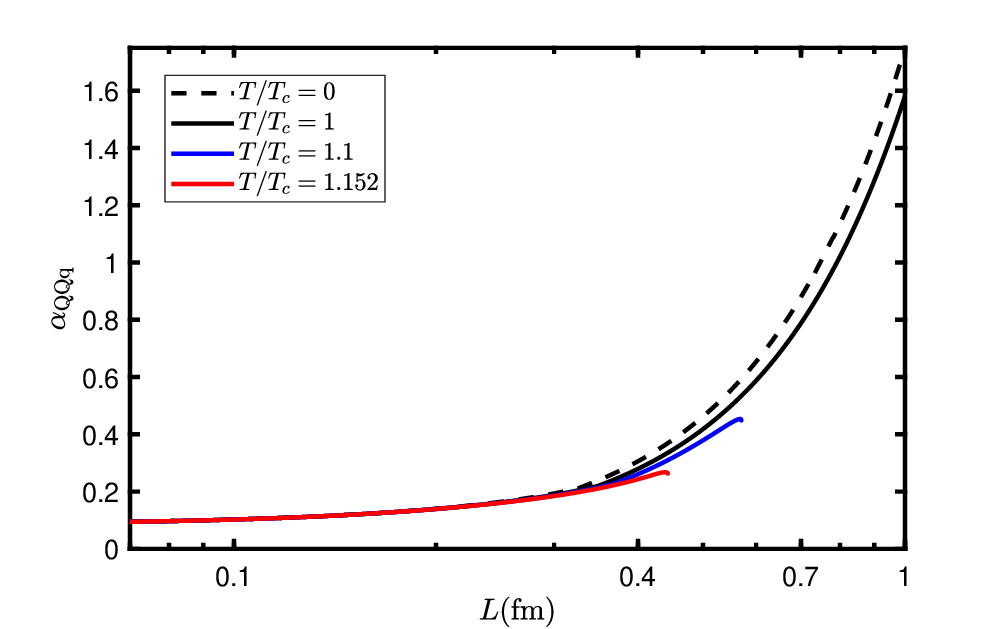}
    \end{minipage}%
    \begin{minipage}{0.45\textwidth}
        \includegraphics[width=7cm]{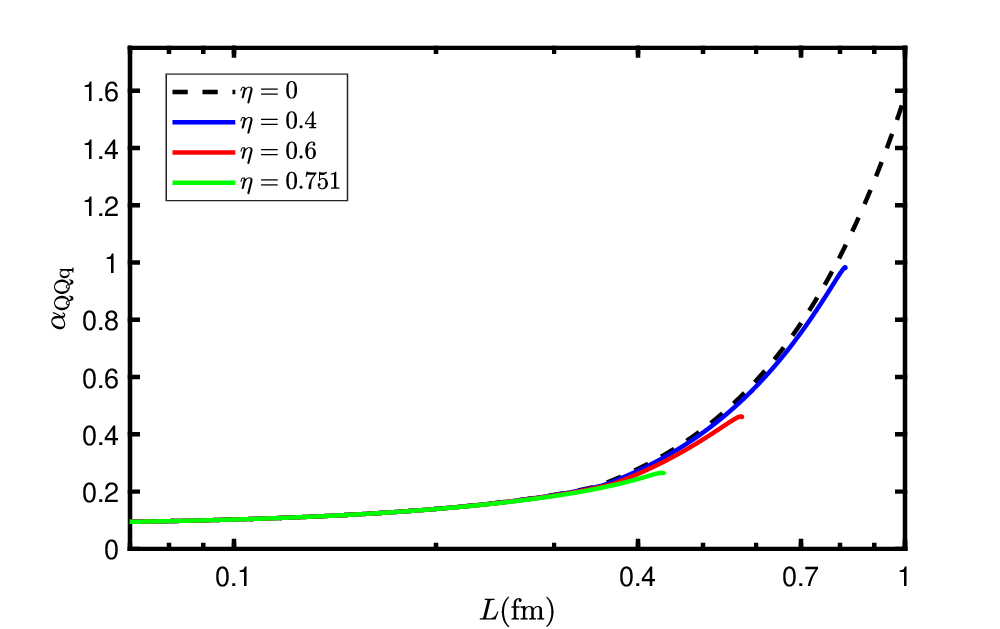}
    \end{minipage}
    \caption{\label{fig12} The left shows the temperature dependence of the effective running coupling for $\mathrm{QQq}$, where $\eta=0, T_c=0.1245\,\mathrm{GeV}$. At $T/T_c=1.152$, it reaches the maximum temperature for $\mathrm{QQq}$ with $\eta=0$. The right shows the rapidity dependence of the effective running coupling for $\mathrm{QQq}$, where $T=0.1245\,\mathrm{GeV}$. At $\eta=0.751$, it reaches the maximum rapidity for $\mathrm{QQq}$ with $T=0.1245\,\mathrm{GeV}$. These plots terminate at the screening distance.}
\end{figure}

As before, we first present the temperature dependence of the effective running coupling for $\mathrm{QQq}$ as shown in the left of Fig. \ref{fig12}, where $\eta=0, T_c=0.1245\,\mathrm{GeV}$. The rapidity dependence of the effective running coupling is shown in the right of Fig. \ref{fig12}, where $T=0.1245\,\mathrm{GeV}$. Their temperature and rapidity values reach up to the maximum values for which $\mathrm{QQq}$ can exist, $T \in [0, 0.1435]\,(\eta=0)$, $\eta \in [0, 0.751]\,(T=0.1245\,\mathrm{GeV})$. As shown in the two figures, the temperature and rapidity dependence of the effective running coupling for $\mathrm{QQq}$ is essentially consistent with that of $\mathrm{Q\Bar{Q}}$. The higher the temperature and rapidity, the lower the effective running coupling curve. Furthermore, the distance at which $\mathrm{QQq}$ reaches maximum coupling is almost identical to the screening distance.

\begin{figure}
    \centering
    \begin{minipage}{0.45\textwidth}
        \includegraphics[width=7cm]{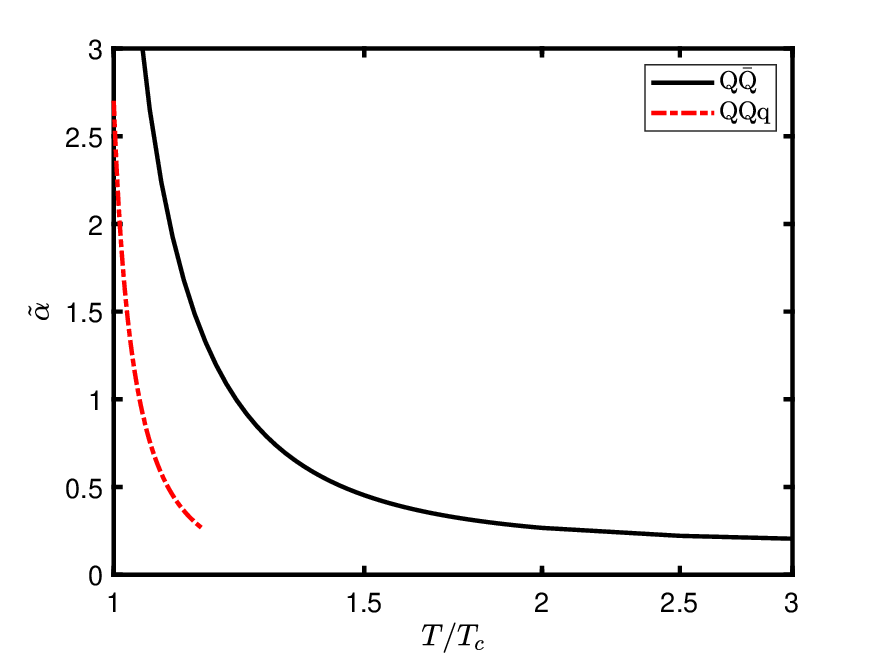}
    \end{minipage}%
    \begin{minipage}{0.45\textwidth}
        \includegraphics[width=7cm]{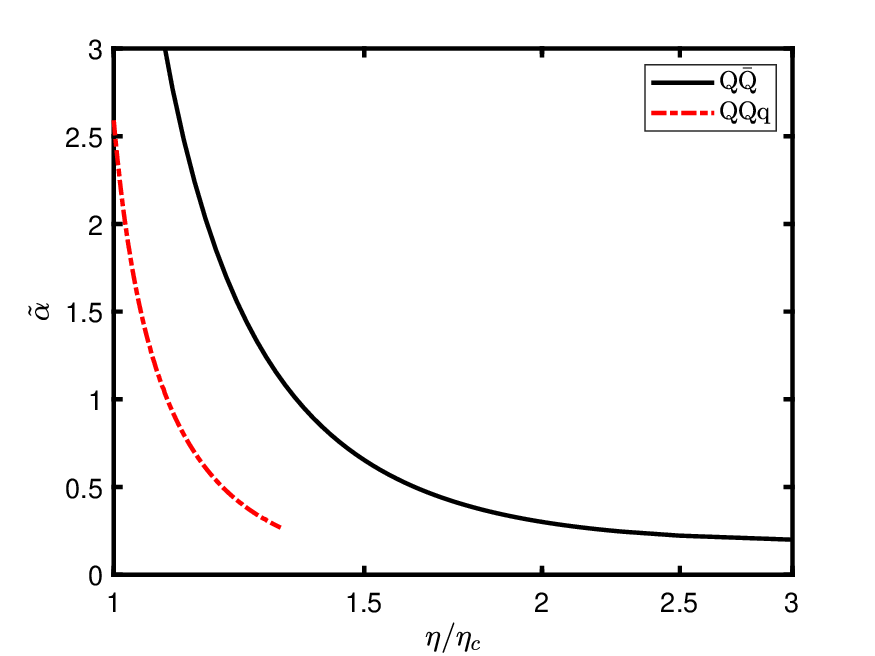}
    \end{minipage}
    \caption{\label{fig13}  The left graph shows the maximum effective coupling constant $\tilde{\alpha}$ as a function of $T/T_c$ at $\eta=0$, where the $T_c=0.1245 \mathrm{GeV}$ for $\mathrm{QQq}$ and $T_c=0.1345 \mathrm{GeV}$ for $\mathrm{Q\Bar{Q}}$. The right graph shows $\tilde{\alpha}$ as a function of $\eta/\eta_c$ at $T=0.1 \mathrm{GeV}$, where the $\eta_c=1$ for $\mathrm{QQq}$ and $\eta_c=1.15$ for $\mathrm{Q\Bar{Q}}$. The black solid line represents $\mathrm{Q\Bar{Q}}$, and the red dotted line represents $\mathrm{QQq}$. }
\end{figure}

\begin{figure}
    \centering
    \begin{minipage}{0.45\textwidth}
        \includegraphics[width=7cm]{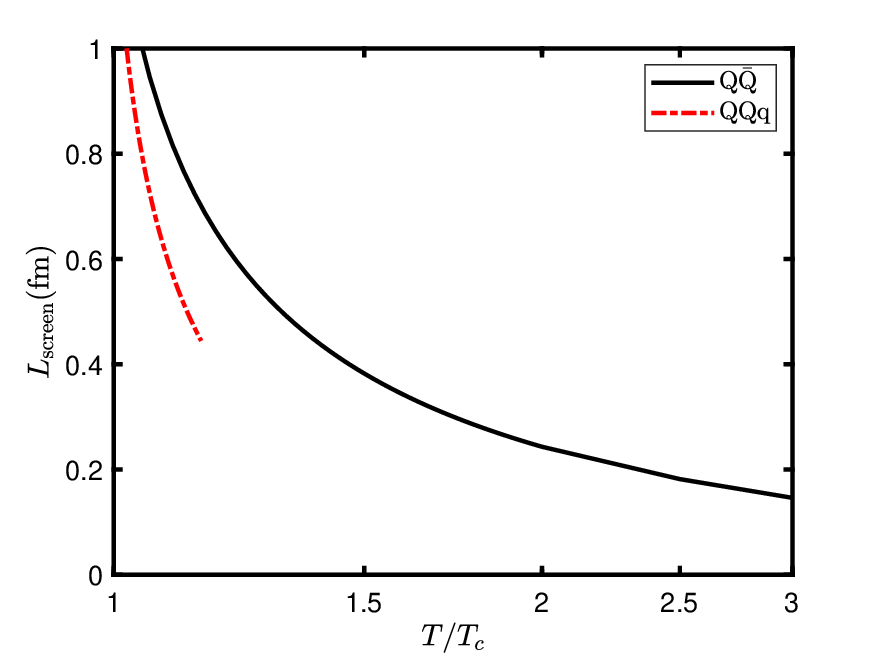}
    \end{minipage}%
    \begin{minipage}{0.45\textwidth}
        \includegraphics[width=7cm]{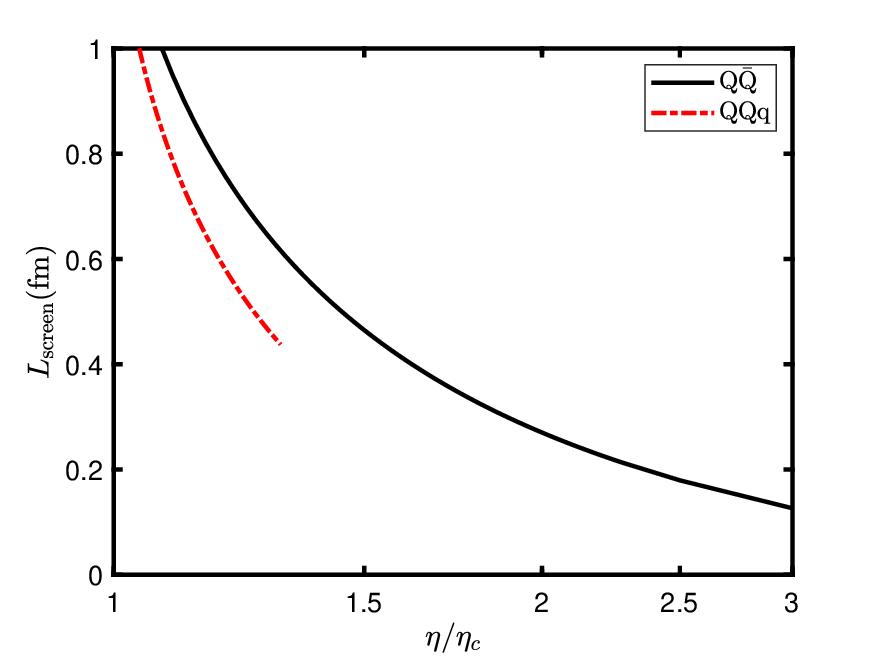}
    \end{minipage}
    \caption{\label{fig14} The left graph shows the screening distance as a function of $T/T_c$ at $\eta=0$, where the $T_c=0.1245 \mathrm{GeV}$ for $\mathrm{QQq}$ and $T_c=0.1345 \mathrm{GeV}$ for $\mathrm{Q\Bar{Q}}$. The right graph shows the screening distance as a function of $\eta/\eta_c$ at $T=0.1 \mathrm{GeV}$, where the $\eta_c=1$ for $\mathrm{QQq}$ and $\eta_c=1.15$ for $\mathrm{Q\Bar{Q}}$. The black solid line represents $\mathrm{Q\Bar{Q}}$, and the red dotted line represents $\mathrm{QQq}$.}
\end{figure}

We focus on the maximum effective coupling constant of $\mathrm{QQq}$, with the maximum effective coupling constant of $\mathrm{QQq}$ in relation to $\mathrm{Q\Bar{Q}}$ displayed as the function of $T/T_c$ and $\eta/\eta_c$ in Fig. \ref{fig13}. And the screening distances for $\mathrm{QQq}$ and $\mathrm{Q\Bar{Q}}$ is shown in Fig. \ref{fig14}. The graphs indicate that the maximum coupling constant of $\mathrm{QQq}$ quickly reduces with an increase in temperature or rapidity, and it consistently remains much lower than that of $\mathrm{Q\Bar{Q}}$, showing a greater sensitivity to both temperature and rapidity. However, $\mathrm{QQq}$ is subject to a limit from the maximum $(T, \eta)$, and its graph ends as the maximum effective coupling constant begins to level off. Similarly, the screening distance for $\mathrm{QQq}$ is invariably shorter than that of $\mathrm{Q\Bar{Q}}$, but it diminishes more quickly.

\section{Summary}\label{sec.4}

In this work, we first show a unified framework capable of accurately describing the behavior of heavy quarkonium and doubly heavy baryons by fitting lattice potentials. Subsequently, we investigate the interaction forces of the two particles in the confined state and at large distances based on the effective string tension. We find that the interaction force of heavy quarkonium is consistently stronger than that of doubly heavy baryons, with its magnitude being approximately twice as large. Furthermore, within this framework, we construct a physically reasonable $T - \eta$ phase diagram. We then examine the interaction forces of the two particles in the deconfined state and at short distances using the effective running coupling. Remarkably, the interaction force of heavy quarkonium remains nearly twice that of doubly heavy baryons, which aligns well with our theoretical expectations.

The coupling constant of $\mathrm{Q\Bar{Q}}$ is extremely low and tends to a small constant within the range of $L < 0.2\,\mathrm{fm}$ or $T > 2T_c$, whereas for $\mathrm{QQq}$, this occurs approximately within $L < 0.4\,\mathrm{fm}$. Due to the force balance equation at the light quark, QQq cannot exist at high energy scales (high temperature or high rapidity). At low energy scales $(T < T_c)$ and large distances, their interaction forces are extremely strong and similarly show almost no temperature dependence. Furthermore, we find that the screening distance of QQq is more sensitive to the energy scale and smaller than that of $\mathrm{Q\Bar{Q}}$ at the same temperature. We believe that this is precisely because the interaction force of QQq is smaller than that of $\mathrm{Q\Bar{Q}}$. Moreover, the critical temperature of QQq is lower than that of $\mathrm{Q\Bar{Q}}$.

Ref.~\cite{Shuryak:2004cy} suggests that a certain number of mesonic bound states in the QGP is one of the key features distinguishing sQGP from wQGP. Since the binding strength can reflect the number of bound states in a system, we can infer that the interaction between heavy quarks in wQGP is significantly weaker than that in sQGP, i.e., the effective coupling is sufficiently small. Following heavy‑ion collisions, an extremely high‑temperature QGP with $T > 2T_c$ is produced, which is in the wQGP regime. Within this phase, the interaction between heavy quarks is extremely weak, and almost no mesonic bound states exist. In this regime, quarks and gluons exhibit nearly independent degrees of freedom, with weak interactions consistent with perturbative QCD predictions.
As the system cools and expands through $T \in [T_c , 2T_c]$, transitioning from wQGP to sQGP where perturbation theory fails, the effective couplings increase rapidly\textemdash a process in which, according to our results, the $\mathrm{Q\bar{Q}}$ interaction is about twice as strong as that of QQq, thereby making mesonic bound states significantly more likely to form.
Since the screening distance of QQq is always smaller than that of $\mathrm{Q\bar{Q}}$, and its critical temperature is lower, heavy quarks preferentially form $\mathrm{Q\bar{Q}}$ mesons during the cooling of the QGP. Only when the temperature drops below the critical value for $\mathrm{QQq}$ do doubly heavy baryons begin to form. The instability of QQq at high energy scales (e.g., due to the breakdown of force balance equations) implies that doubly heavy baryons are unlikely to exist under extreme temperatures or rapidities. This may explain the relatively low production rates of particles like $\Xi_{cc}^{++}$ in LHC experiments. Increasing rapidity significantly reduces the screening distances and maximum coupling constants of both $\mathrm{Q\bar{Q}}$ and QQq, with the effect being most pronounced in $T \in [T_c , 2T_c]$. By comparing the effective couplings at different rapidities, we find that QQq is more sensitive to rapidity than $\mathrm{Q\bar{Q}}$. This difference may arise from the dynamical effects of light quarks, offering new perspectives for studying energy loss mechanisms of light quarks in the QGP.

\section*{Acknowledgments}
This work is supported by the NSFC under Grant Nos. 12405154, Open Fund for Key Laboratories of the Ministry of Education under Grants Nos. QLPL2024P01, the European Union -- Next Generation EU through the research grant number P2022Z4P4B ``SOPHYA - Sustainable Optimised PHYsics Algorithms: fundamental physics to build an advanced society'' under the program PRIN 2022 PNRR of the Italian Ministero dell'Universit\`a e Ricerca (MUR).

\section*{References}
\bibliography{ref}
\end{document}